\documentclass{elsart5p}
\usepackage{natbib}

\usepackage{amsmath,amssymb}
\usepackage{graphicx,psfrag,units}

\makeatletter
\def\ps@copyright{\let\@mkboth\@gobbletwo
  \def\@oddhead{}%
  \let\@evenhead\@oddhead
  \def\@oddfoot{\small\slshape
    \def\@tempa{0}
    \ifx\@volume\@tempa
      \hfil\@date\/%
    \else
      Article published in \@jou@vol@pag\hfil\hbox{}\fi}%
  \let\@evenfoot\@oddfoot
}
\def\paragraph{\@startsection{paragraph}{4}{\z@}{2ex \@plus
  0.5ex \@minus 0.2ex}{-1em}{\normalfont\normalsize\itshape}}
\makeatother

\begin{document}

\begin{frontmatter}
\journal{J. Math.\ Psych.}
\title{Grandmother cells and the storage capacity of the human brain}
\author{John Collins}
and
\author{Dezhe Z. Jin}
\address{Physics Department,
   Pennsylvania State University,
   University Park, PA 16802, USA
}
\date{26 February 2007}

\begin{abstract}
  Quian Quiroga et al.\ [Nature \textbf{435}, 1102 (2005)] have
  recently discovered neurons that appear 
  to have the characteristics of grandmother (GM) cells.  Here we
  quantitatively assess the compatibility of their data with the GM-cell
  hypothesis.  We show that, contrary
  to the general impression, a GM-cell representation can be
  information-theoretically efficient, but that it must be accompanied
  by cells giving a distributed coding of the input.  We
  present a general method to deduce the sparsity distribution of the whole
  neuronal population from a sample, and use it to show there are two
  populations of cells: a distributed-code population of less than
  about 5\% of the cells, and a much more sparsely responding
  population of putative GM cells.  With an allowance for the number
  of undetected silent cells, we find that the putative GM cells can
  code for $10^5$ or more categories, sufficient for them to be
  classic GM cells, or to be GM-like cells coding for memories.  We
  quantify the strong biases against detection of GM cells, and show
  consistency of our results with previous measurements that find only
  distributed coding.  We discuss the consequences for the
  architecture of neural systems and synaptic connectivity, and for
  the statistics of neural firing.
\end{abstract}
  
\end{frontmatter}

\maketitle

\section{Introduction}

Critical to understanding how information is processed in the brain is
the form of the  neural coding that underlies the storage and
recall of memories.  Is there a local, or gnostic \citep{Konorski},
code --- colloquially called a grandmother-cell (GM-cell)
representation --- in which the firing of a single neuron (or group of
neurons) exclusively codes recognition of a particular object, person
or memory?  Or is the code much more distributed?

Although it is generally accepted [e.g., \cite{Churchland.Sejnowski}]
that GM representations are not used in reality, experiments
\citep{HVC-RA,dg.cell.classes,silent.cells} often find localist
responses by individual neurons.  Most dramatically, 
\citet{GMC} have recently found many neurons in humans that,
within the limits of the measurements, behave like classic GM cells.

In this paper, we therefore quantitatively re-examine the viability of
GM-cell representations, with the outcome that we refute the standard
quantitative arguments against them, both theoretical and
phenomenological.  The information-theoretic argument is that GM
representations need far too many neurons for the information coded
\citep{Rolls.Treves,Rolls,Churchland.Sejnowski}.  We show that this
argument fails when one examines the information storage capacity of
the synapses rather than the representational capacity of neurons for
input stimuli.  The standard efficiency argument applies only to the
input representation, needed to represent any of the myriad possible
stimuli.  For storage, a GM representation can be optimally efficient.

The phenomenological argument is that GM cells should fire in response
to a much smaller fraction of stimuli than has been deduced from
measurements of neural responses \citep{HKPC,Abbott.Rolls.Tovee,Waydo}.
A GM cell can be regarded as a categorizer, and the data appear to
imply that any apparent GM cell responds to many categories of stimuli
rather than to one category.

However, our information theoretic argument shows that associated with
any GM-cell population, with its ultra-low sparsity, is a more
conventional population with a much higher sparsity.  This
two-population property, always a part of the GM-cell idea
\citep{Konorski,Page}, was not allowed for in older analyses, including
that \citep{Waydo} by the group responsible for the new data
\citep{GMC}. We devise a very general method of analyzing neural
systems with multiple sparsities, and apply it to the data of
\citet{GMC}.  It enables us to quantify the biases against experimental
detection of GM-like cells, most of which simply appear as unreported
silent cells, and whose estimated numbers \citep{Waydo,Henze,Buzsaki}
--- may be a factor of 30 more than the reported cells.

We find that the two-population property holds, and that less than
$5\%$ of \emph{detected} cells are in the distributed-code population:
the vast majority of the cells can be GM-like.  Then we find that the
number of categories coded by the GM-like cells can be $10^5$ or more.
Uncertainties are minor relative to the orders of magnitude involved.
The data of \citet{GMC} therefore appear in strong quantitative
agreement with the GM-cell hypothesis.  

The biases against detecting GM cells are enough to allow consistency
with previous \citep{Abbott.Rolls.Tovee,Waydo} measurements and
analyses that use a single-population model and that quantitatively
argued against GM cells. We will examine other arguments against GM
representations in the Discussion section.

\section{GM systems}

Inappropriate or excessively rigid definitions can exclude
biologically interesting cases.  For example, 
\citet[p.\ 12]{Rolls.Treves} define a local representation as one where
``all the information that a particular stimulus or event occurred is
provided by the activity of one of the neurons''.  The word ``all''
appears to exclude a system where a local representation codes the
result of testing a distributed representation of a stimulus against
remembered items.  Not all the available information is in the firing
of the neurons in the local output representation.

Therefore in this section we present our definitions, and explain
important features and consequences of the definitions needed for
later sections.

\subsection{Definitions}

We define a set of cells to form a local or GM-like system when nodes
of the system can be divided into groups of one or more nodes, and, to
a good approximation, each group corresponds to one particular
meaningful and distinct property of the stimulus input to the system.
Each group we call a GM group.  Typically we treat the properties
corresponding to different GM groups as being mutually exclusive.  We
will usually identify the nodes with actual neurons, so that
measurement in one of a GM group's cells of firing above a suitable
threshold is strong evidence that the stimulus is associated with the
corresponding property.  But it is also possible that the nodes could
be, for example, part of a dendritic tree.  Then the correspondence
between neural firing and local coding might not be direct.  In any
case we can treat the system as a categorization system.

In contrast, a distributed representation is formed by a set of cells
where the categorization can only be determined from the activity of
multiple cells/nodes and where the patterns of activity overlap
between distinct properties, even when the properties themselves are
mutually exclusive.

\paragraph*{Classic GM cells and generalizations}
The classic example of GM system is a facial identification system,
where the firing of a particular GM group mediates the ``unitary
perception'' \citep{Konorski} of a retinal image as corresponding to a
particular individual person.  A characteristic property of a
\emph{classic} GM system is therefore that firing is exclusive between
different GM groups --- i.e., only one GM group is active at a time.
This corresponds to the fact that the individuals associated with the
different stimuli are themselves completely distinct.  Of course,
there will be situations where the GM group firing is not completely
exclusive; for example, if a particular stimulus is ambiguous, or if
the picture of a face of one person is artificially morphed into that
of another person.  

Our definition, however, was worded to allow certain natural
generalizations from the case of classic GM cells.  In particular, we
will apply the terminology to declarative memories in general
(episodes, facts, etc).  Thus we could have a GM cell or GM group
corresponding to each episodic memory.  In this case it is evidently
of practical importance for a memory to be recalled from a stimulus
containing a few components of the original memory.  Since the same
components could be part of other memories, the pattern of recognition
firing need not be exclusive between memories.  Let us regard these
patterns as priming the recall of the memories.  Full conscious recall
of one particular memory requires some extra cues and modulation.
With a GM-like memory system, non-exclusive priming recall would be at
a relatively low level of firing above some threshold, with full
recall involving exclusive firing at a much higher level.

In this case the exclusivity is not between the actual firing of
different GM groups, but between the concepts corresponding to the
groups.  Of course, the situation is a little more complicated for
episodic memory, since episodes are happenings along a continuum in
time.  A memory cell for an episode corresponds to a small range in
time.  The exclusivity between different GM groups is between
well-separated episodes, of which there are evidently a very large
number.  

Another natural generalization of the GM-cell concept is to local
coding for output, with strong experimental evidence in the work of
\citet{HVC-RA}.

\paragraph*{High-level GM systems v.\ low-level local coding}
We choose to restrict our use of the ``GM-cell'' terminology to the
higher levels of neural processing.  For lower levels, we will use the
broader term ``local coding''; by this we mean coding by a set of
cells each of which is responsive to some patch of stimulus space
(with fall off at the edges naturally).  When the relevant properties
of the stimulus are in a low-dimensional space --- as for color or
position in an environment --- the collection of patches can give
complete coverage.  But when the relevant stimulus space is high
dimensional, one can only expect local coverage of a minute fraction
of possible stimuli --- for example, to correspond to linguistic
phonemes out of all possible auditory stimuli.

There is considerable evidence that can be interpreted as supporting
some kind of local coding below the highest perceptual levels.  The
key question for us is whether local coding is also used at the
highest levels.

\paragraph*{Internal stimuli}
Input to a memory system need not correspond to actual physical events
external to the brain.  Some input can be completely generated
internally, as with an author planning a novel.  Memories of people
and episodes in the novel have the same neural status as memories of
real people and events.

\subsection{Properties}

\begin{figure}
  \centering \includegraphics[scale=0.85]{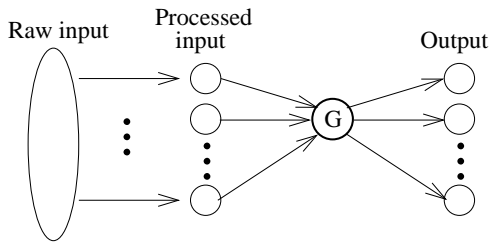}
  \caption{Cell G is a classic GM cell that fires in response to a
    certain kind of visual pattern (e.g., the face of a particular
    person).  It receives processed input from the top of a hierarchy
    that processes raw sensory input.  The GM cell passes recognition
    information to a set of output cells, which could, for example,
    recall the name of a recognized person.}
  \label{fig:gm-struct}
\end{figure}

One big computational advantage of a GM-like system is the simplicity
of information flow, Fig.\ \ref{fig:gm-struct}, so that the
representation is easy to construct and manipulate --- e.g.,
\citet{GMB,BMW}.  New memories are formed with new or unallocated
neurons, so that they interfere minimally with old memories
\citep{Quartz.Sejnowski}.  The input to GM cells is from a distributed
representation of the stimuli.  Output to downstream neurons using the
categorization can be simply taken from a single GM cell.

Above-threshold firing of a GM-cell models a property of the cause of
the stimulus.  Thus when firing is exclusive between different GM
groups, this corresponds to exclusivity between the modeled 
properties of the associated external stimuli, like the
identity of a person.

It is therefore tempting to use exclusivity of firing between
different GM groups as the primary measurable criterion for
characterizing GM-like systems.  But natural generalizations to
declarative memory motivate us to relax this criterion.  For example,
with episodic memories, a stimulus may cause a response in nodes for
those episodes having important commonalities with the stimulus.

In addition, there can be multiple categorization systems, and there
is no requirement of exclusivity between different systems.  This is
particularly clear at low levels in the processing hierarchy, where we
could have separate local representations, for example, of the color
and shape of an object.  A collection of local representations of such
relatively low-level features then forms a distributed representation
of the whole object, suitable for input to the next level of
hierarchical processing.

Note that we prefer to use the terminology ``GM-like system'' rather
than ``GM representation'' to emphasize two aspects: The first is that
if the GM-cell idea applies in its classic sense of recognition of
individual people, it is likely to apply much more generally to all
declarative memories.

The second is that we wish to treat the firing of a GM cell as coding
the result of a \emph{recognition computation} from a stimulus.  But
the use of the word ``representation'' for a GM-like system would
carry the connotation of representing the stimulus, which is not
generally appropriate.  If nothing else, local coding typically
dramatically fails to cover the stimulus space.  For example, consider
a distributed input representation on 100 binary neurons, a small
number compared with real sensory systems.  There are $2^{100} \simeq
10^{30}$ distinct stimuli, many orders of magnitude larger than the
total number of neurons in any brain.  A practical local
representation can only apply to a minute fraction of stimuli,
presumably ones that are especially salient.  A local representation
can only provide full coverage for a stimulus space of very low
dimension, like that for color.

\subsection{Detection of GM-like systems}

Practical experiments only involve a limited number of stimuli and
cells, and definitely do not give the detailed synaptic information
that determines all possible causes of a cell's firing.  Thus it is
non-trivial to distinguish a GM system from a distributed
representation, when the distributed representation is sparse, and
when we allow natural generalizations of the GM-cell idea.

We illustrate the issues by comparing two very different computational
memory models, one by \citet{Hopfield}, and one by
\citet{BMW} (BMW).

\paragraph*{Hopfield's model}
In Hopfield's particular example, the input stimulus concerns
properties of people, and the input representation is carried by a set
of 1000 binary neurons.  These are divided into 50 sets of 20 neurons.
The 20 neurons in each set give a local representation of 20 possible
values of a property of the stimulus.  For example, to input the name
of a person, one of a set of 20 name-coding neurons would be active.
Binary synapses connect every neuron to every other neuron.

Each stored memory is considered as the set of 50 property values for
a particular person, and is coded in the state of the synapses.  The
synaptic strengths are set by a Hebbian-like rule on presentation of
stimuli.

Recall of a memory is caused by a stimulus that consists of partial
data about an individual, i.e., values for a subset of the 50
properties.  Memory retrieval results in completion of a partial
stimulus to the full set of properties for the corresponding
individual.  There is no corresponding GM cell; the model is a fully
distributed memory system.

\begin{figure}
  \centering
  \includegraphics[scale=0.6]{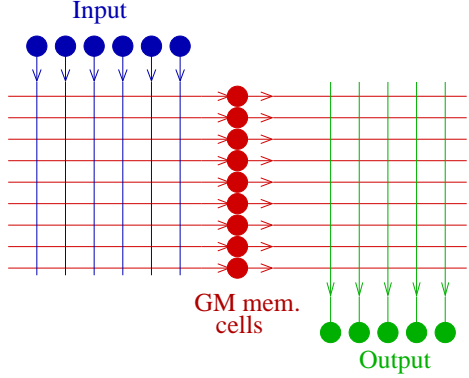}
  \caption{Architecture of BMW model in its most basic form.}
  \label{fig:BMW}
\end{figure}

\paragraph*{BMW model}
The basic BMW model is the simplest possible form of a GM system: a
feed-forward perceptron with one intermediate memory-cell layer, Fig.\
\ref{fig:BMW}.  The memory cells are arranged to respond in a GM-cell
style to recognize inputs that correspond to stored memories.  The
model gains its power by applying it to the situation that the input
forms a sparse binary representation of stimuli.  This would typically
be obtained from the top of a processing hierarchy, as in Fig.\
\ref{fig:gm-struct}.  Improvements in the model can be made, for
example, by adding inhibitory interneurons to enforce a
winner-take-all action in the memory-cell layer, but the simplest form
of the model is sufficiently robust for our illustration.

To solve the same pattern-completion task as Hopfield's model, we make
the output cells identical to the input cells, thereby specializing
from a heteroassociative memory system to a homoassociative system.

Memories are created by presenting a full stimulus to the system, and
arranging for some unallocated memory neuron $\alpha$ to match its synaptic
strengths to the stimulus.  The chosen neuron changes to a state of
being allocated, with correspondingly greatly reduced synaptic
plasticity.  Let $\mathbf{x}^{(\alpha)} = \big( x^{(\alpha)}_1, \ldots,
x^{(\alpha)}_{1000} \big)$ denote the prototype pattern for memory neuron
$\alpha$.  Then the strengths of the neuron's input synapses are set to a
constant times the input pattern: $w_\alpha{}^j = Wx^{(\alpha)}_j$, and
similarly for the output.  The constant $W$ can be scaled out of all
of our formulae, but its use assists in relating the formulae to
properties of real neurons.  When another stimulus $\mathbf{x}'$ is
presented, the response of neuron $\alpha$ is obtained from a thresholded
sum over its inputs:
\begin{equation}
  r_\alpha = \theta\Big( \sum_j w_\alpha{}^j \, x'_j - T_\alpha \Big).
\end{equation}
This formulation is for artificial analog neurons, but it is readily
generalized to realistic spiking neurons, and the model is robust
enough that minor variations do not affect the principles governing
its performance.

We now review how standard properties \cite{kanerva}
of sparse binary codes show that the
system implements the pattern completion task if the threshold is set
suitably.  Let a stimulus $\mathbf{x}'$ be presented; it may be some
full pattern, in which $5\%$ of the input neurons are active, or it may
be a partial pattern.  The input to neuron $\alpha$ is $\sum_j w_\alpha{}^j \,
x'_j$, which is $W$ times the overlap between $\mathbf{x}'$ and the
prototype pattern for memory $\alpha$, i.e., it measures the number of
common on-bits.

If the new stimulus is unrelated to memory $\alpha$, then its on-bits are
random relative to those for memory $\alpha$.  Thus for a full pattern the
typical overlap is $5\%$ of the total number of on-bits in a full
patten, i.e., $5\% \times 50 = 2.5$, and less for a partial pattern.

We therefore set the threshold intermediate between the input for the
full prototype pattern, i.e., $50W$, and the typical input for a full
unrelated pattern, i.e., $2.5W$.  Then unrelated patterns almost never
cause neuron $\alpha$ to fire.  Thus with sparse input patterns we have the
well-known automatic orthogonalization against unrelated stimuli.

But now consider an input corresponding to a subset of the features in
memory $\alpha$'s prototype pattern.  For example, a quarter of the
features could have been detected.  In that case, the input to neuron
$\alpha$ would be $\frac14\times 50W=12.5W$.

Hence with an appropriate threshold setting, we get firing of neuron
$\alpha$ in response to a subset of the features in its prototype pattern.
This then causes firing of the whole set of output neurons that
correspond to the prototype, i.e., we have pattern completion.

Because the input to a memory cell is much bigger for patterns related
to its prototype than for unrelated patterns, the operation of the
system is robust against changes in the exact equations describing its
operation.

\paragraph*{Measurements}
To mimic a biological experiment, one could measure the responses of a
sample of model neurons to a sample of stimuli.  In Hopfield's model
if there are fewer than about 10 or 20 stimuli, the firing of many of
the individual neurons would correspond to single people.  Thus the
distributed nature of the memory representation would not be
immediately apparent.

But it is not necessary to enlarge the stimulus set to test this.
Data from a sample of neurons and stimuli give the fraction of stimuli
to which neurons respond. Then by using knowledge of the capacity of
the system, which in this case is several hundred memories, one can
statistically extrapolate from the sample to show that each neuron is
responsive to multiple unrelated stimuli.

As for the BMW model, the input/output cells would have the same kind
response characteristics as all the cells of the Hopfield model,
responding to $5\%$ of the stimuli.  But there are also GM memory
cells, which respond much more rarely.  If the sample data include at
least a few responsive GM neurons, one can detect the different
properties of the input and GM cells, and thereby distinguish the
system from the Hopfield model.

In this paper, we will construct a method to extrapolate in general
from data with a sample of cells and stimuli to the whole system.  The
method enables us to compute which of the following properties of a
set of putative GM cells is consistent with data: (a) The cells
actually could code for single properties. (b) Each cell is expected
to respond to multiple \emph{unrelated} properties, after an
extrapolation to the full set of possible stimuli.

One confounding issue is that if one detects a response from a GM-like
cell it is easy to misidentify the property corresponding to the
cell's firing.  If a cell only responds, within limited data, to a
particular individual person, the cell could indeed be a classic GM
cell corresponding to that person.  But it could also be, for example,
a GM cell for an episodic memory that contains the person.  In that
case it would respond to stimuli containing other components of the
episode. 

The distinguishing feature of the most general kind of GM-like cell,
but one that can be hard to test, is that apparently different stimuli
that cause it to respond are in fact related.  In contrast, with a
purely distributed representation at the highest level, there is no
high-level relation between, for example, the multiple people causing
a particular cell to respond.  The only relation is an identity of
lower-level features in the input representation.  The locality is at
the feature level, not at the high level.

\section{Information requirements for pattern recognition}
\label{sec:info}

We now quantify the information requirements for a memory or
categorization system.  First we distinguish the activity state and
the storage state.  The activity state is the pattern of firing of the
neurons, and the storage state \citep[p.\ 142]{Churchland.Sejnowski} is
the pattern of synaptic connectivity and strengths.

Furthermore, within the activity state of a memory system, we
distinguish an input representation and a recognition representation.
The input representation is of the current input stimulus, like a
visual scene, while the recognition representation concerns which
stored pattern (if any) corresponds to the current input.

\subsection{Input representation}

The immediate input to a memory system must be able to represent the
relevant features of any possible stimulus, and not just those
previously encountered stimuli for which there is a memory trace.
Here the standard arguments for distributed representations apply
unambiguously, and a GM representation is not possible.  The argument
is simply that $N$ neurons code for at most $N$ exclusive properties
in a GM system, but that they code for exponentially more with a
distributed representation, for example $2^N$ with a simple binary
code.

In general the immediate input to a memory or categorization system is
not the raw sensory input but a highly processed representation of
those high-level features that are relevant for the system's
particular task.  For example, input for face recognition could
involve neurons coding for the presence, shape and position of eyes,
nose and mouth, etc; individual features could well be coded locally,
but the collection of feature representations would always form a
distributed representation.  Each neuron has a range of
distinguishable firing rates, so that the raw information capacity in
the activity of $N$ neurons is a few times $N$ bits.  But robustness
requires a certain amount of redundancy, and firing is often sparse,
both of which reduce the information per neuron.  Measurements
\citep{Abbott.Rolls.Tovee} show that in hippocampal neurons about $0.3$
bits of independent information are coded per neuron and suggest that
a few hundred neurons suffice for coding possible faces.  For example,
300 neurons code about 100 bits, sufficient for about
$2^{100}\simeq10^{30}$ different categories of face, an entirely
satisfactory number.

Note that once an input representation is sufficiently small, pure
representation efficiency is not a dominant consideration.  Issues of
processing speed, metabolic efficiency, and algorithmic robustness can
be more important.  For example, sparse distributed representations
appear to be favored \citep{Rolls.Treves}, since they can give weak
interference between memories during synaptic plasticity.

\subsection{Storage in synapses}

The information capacity in the storage state, i.e., in the synapses,
was estimated by \citet{CMS} and by
\citet{SHC}.  There are two contributions, from the synaptic
topology and from the synaptic strengths, giving a total of 5 to 10
bits of raw storage capacity per synapse.  We divide by 10, to provide
a plausible allowance for redundancy.  Thus we need about one synapse
for each bit of storage information.

This calculation uses only very basic physical information about
synapses and neural processing, so it is certainly accurate at the
order of magnitude level. Since measures of information in units of
bits are independent of the physical implementation, the numbers are
directly compared with those for ordinary digital computers, and
experience with data processing and storage can be used to derive
minimum numbers of synapses for a task.  A human brain
has around $N_{\rm tot} \simeq 10^{11}$ neurons and $C\simeq10^4$ synapses per
neuron.  So its synapses store about $CN_{\rm tot} \simeq 10^{15}$ bits,
i.e., about $\unit[10^5]{GByte}$, up to a factor of 10 or so.

\subsection{Recognition representation and total stored information}

Suppose the system has a repertoire of $R$ stored memories.  Each is
an arbitrary association of a category to stimulus features.  So we
attribute to each memory $A$ bits of association information, which we
term its semantics.  This includes both input and output information.
(E.g., facial structure and name for a person, and, in fact, \emph{all
  remembered information about the person}.)  It is important to
include output semantics in the associations, since they are what
allow the retrieval of a particular memory to cause memory-specific
behavior.  These are quite arbitrary associations, for example of a
linguistic name to a specific person, and without such output
associations there is no purpose to storing a memory.  The output
associations do need to be stored, and they therefore contribute to
the calculation of the minimum size of the system just like the input
associations.

The total association information is $RA$ bits, so that approximately
this number of synapses is needed.  For example, with a repertoire of
$R=5000$ and an average of $A=10000$ bits of information per item,
we need about $5\times10^7$ synapses.  The measured number of synapses
per neuron is around $10^4$, thereby implicating about $RA/10^4$
neurons, i.e., a number proportional to the size of the system's
repertoire (and the information per memory trace).  

Hence if the system uses $R$ GM cells for a recognition
representation, this is at most a constant factor beyond the neurons
needed to carry the storage synapses.  Moreover, if $A$ is larger than
about $10^4$, there is not even any overhead at all in using GM cells.
We can characterize this by saying that if the memories are
semantically rich, then a GM strategy for the recognition output can
be optimally efficient, as in the model of 
\citet{BMW}. 

In contrast, if one ignored the storage requirement, one would assert
that $N$ recognition neurons can code exponentially more recognized 
categories, e.g., $2^N$.

It is important that in quantifying the information the term ``bit''
is used in the strict information theoretic sense.  This means that if
each memory were coded as an actual bit pattern, each of the $2^A$
possible patterns would be equiprobable.  Thus, on a computer, the bit
count refers to an optimally compressed representation.  However, only
certain features of the input are relevant for tasks like face
identification: a minimalist line drawing often suffices for
unambiguous identification.  So in computer terms the association bits
are with respect to a representation that is both lossy and
compressed.  In the opposite direction, it can be difficult to perform
computations on optimally compressed representations, and it is also
difficult to measure accurately the probabilities of occurrences of
different kinds of stimuli, since the number of possible stimuli far
exceeds the number actually experienced.  Moreover redundancy (in the
information theoretic sense of using more than the minimum necessary
number of bits) is useful in giving robustness to a system.  For all
these reasons, we must expect the physical capacity of a system needed
to code memories to be a substantial factor large than a minimal
physical implementation of $RA$ bits. Nevertheless this measure is
important in quantifying information in an implementation-independent
way.  It also enables us to estimate the information storage
requirements by examining implementations of related tasks on a
digital computer.

Well-known examples of simple line drawings and pictures of
artificially low resolution show that the information to identify
faces could be quite modest, if a suitable representation is used.
But the synaptic size of the system also depends on the remaining
association information, treated as output.  This can be much larger
in size, effectively amounting to a biography of the individual
concerned in each memory.

\subsection{Objections and answers}

A number of objections to our bounds and ideas for evading them have
been proposed, which we now answer.  The general answer is simply that
the information theoretic bound represents an absolute physical limit
that it is impossible to exceed.  All that is required is that we
count the bits of information in the strictly correct information
theoretic sense, and that we have identified the correct physical
location of memory in the synapses.

\paragraph*{Counting memories} Does not the idea of quantifying
memories as discrete items that can be counted carry the implication
that we use a GM system?  Are there not difficulties in counting
memories in distributed memory systems?  In fact the classical kinds
of distributed memory, e.g.,
\cite{Hopfield,Treves.Rolls1,Treves.Rolls2,Amit.Brunel.Tsodyks}, are
regarded as storing patterns, which can be counted.  What we do have
in mind is declarative, or explicit memory, the kind considered as
prototypically hippocampal.  Here it is reasonably clear what is meant
by a single memory: a picture, a scene, or the meaning of a word; all
of these are discrete.  Even with episodic memory, where there is a
continuous time variable, we can observe that there is a correlation
time within a continuous series of events.  As regards storage
requirements, we simply identify a single memory with the happenings
within a correlation time, which is evidently of the order of seconds
or minutes.  Of course, only a small fraction of these are stored in
long term memory.  We will not require great precision here.

In the contrasting case of implicit procedural memory, it is much less
obvious what should be defined as a single memory item.  But we are
not concerned with this case.

\paragraph*{Multiplexing}
Could not one gain by allowing a neuron to respond to multiple
different stimuli?  Could not a single face-identification cell
respond to either George Bush, Jennifer Aniston or John Hopfield, for
example?  This multiplexing is just going in the direction of a
distributed representation.  Our argument so far does not rule that
out; all it says is that this does not provide a way of reducing the
synapse count.  It is our statistical argument in later sections that
enables us to estimate the degree of multiplexing.

Now if a neuron responds to multiple categories, then there is
interference at the neural level between different memory traces.  An
unambiguous categorization then requires the use of the firing
information from more than one neuron.  In the case of lightweight
memories, i.e., with substantially less than 10000 bits of
associations per memory, multiplexing of memories can indeed reduce
the neuron count, and is allowed by our general argument.  But with
richer memories, there is no gain.

\paragraph*{Representation v.\ memory}
The memories we have discussed are typified by hippocampal memories.
Consider instead visual area V1; the number $P$ of distinguishable
activation patterns is exponential in the number $N$ of neurons, e.g.,
$P=2^N$.  An outside observer could identify different faces from the
different activation patterns.  Why should we not regard this as
memory, if we are to regard observation of (much simpler) activation
patterns in GM cells as identifying faces, and as in fact part of a
memory system?

The difference is not in the outside observations, but in the use the
organism itself makes of the information.  A memory is not useful
unless it produces some consequence when recalled.  What we mean, for
example, is that seeing John Hopfield on the other side of a street
might induce us to cross the street and greet him by name.  For the
billions of other possible people, there would be either be a
different response or no response at all.

This is why we defined the associations for a memory to include output
as well as input.  There must be sufficient information stored to
enable the relevant computation to be done from the activation pattern
corresponding to the stimulus.  

Without being concerned with storage, it is perfectly possible to
compare one activation pattern with a previous one, to identify
whether or not it has changed.  But to compare it with patterns of
activation for all the people one remembers, and to take appropriate
actions, one needs appropriate storage.  It is simply not possible to
evade the fundamental physical necessities given by information
theory.

\paragraph*{Coding commonalities}
Many remembered faces have features in common.  Cannot this be used to
reduce the number of synapses and neurons needed, by coding the common
features in a special subsystem for the common features?  Suppose we
had a set of faces characterized by very short, dark, and curly hair.
Would there not be a gain by allocating a neuron to this combination
of characteristics and using it instead of separate neurons for hair
length, color and curliness?

If the different hair features were equiprobable and uncorrelated, the
general argument would prevent any gain.  As an example, suppose each
of the three hair characteristics has 8 equiprobable values, for
$8^3=512$ distinguishable combinations.  If all the combinations were
equiprobable, we would allocate 9 bits for the information content.
Probability here refers to a prior probability of occurrence, i.e.,
before the creation of a memory trace.

But if instead all the characteristics were perfectly correlated, then
there would only be 8 combinations, which could be represented in 3
bits.  When we construct memories, this gives a gain of a factor of 3
in storage if we only represent the combinations that actually occur
in the input representation rather than all possible combinations.

But this is exactly what is meant by using a correct measure of
information to compute the minimum number of synapses.  Note that the
task carried out by a memory system is not merely to identify the best
fit to a current stimulus among stored memories; for that very few
bits are needed.  For example, if a stimulus is represented by 100
bits, but only 8 memories are stored, then only $\log_2 8 = 3$
suitably coded bits are needed to identify the stimulus, if it is
assumed that the stimulus corresponds to one of the memories.  But it
is also necessary to identify the case that the stimulus fails to
correspond to a stored memory, so that it is a candidate for a new
memory.  It is for this that the other 97 bits are needed.

\subsection{Comparison with models}

Although our argument was used to show that GM coding can be optimally
efficient in the use of synapses and neurons, the bounds on synapse
and neuron number are independent of the coding method.  It is
therefore useful to verify that the bounds are obeyed by the polar
opposite of GM systems, i.e., by conventional distributed memory
systems.  Calculations of the capacity of such systems have been made,
e.g., \cite{Treves.Rolls1,Treves.Rolls2,Amit.Brunel.Tsodyks}, and it
can be checked that they do obey our bounds.  But what appears to have
been missed is that the capacity limit also removes the argument
against using GM cells.

Hopfield's recent model \citep{Hopfield} provides an excellent example.
In this model, each stored memory corresponds to the values of each of
50 categories, with 20 possible values per category.  This gives a
total of $50 \log_2 20 = 216$ bits of information.  Retrieval of a
memory results in a neural representation of this information in the
firing of 50 out of 1000 binary neurons, for a biologically realistic
sparsity of $5\%$.

Storage is in binary synapses with all-to-all connectivity on the 1000
neurons, for a total of slightly under $10^6$ bits of storage
capacity.  This implies that the system can store at most $10^6/216 \simeq
4500$ separate memories.  In fact, the simulations in \citet{Hopfield}
show that, with the algorithms used in that paper, performance
noticeably degrades when about 250 memories, i.e., a factor of 20
below the physical limit.  Thus the information-theoretic bounds are
obeyed.  The problem is that the memories are stored on the synapses
connecting the active neurons in a particular memory.  Synapses
overlap between memory traces, which causes interference if too many
memories are stored.

Consider in contrast the BMW model \citep{BMW}, which, as observed by
its authors, is optimal in its number of synapses.  Suppose for input
and output we use the same 1000 neurons as in the distributed model.
Then to store $N$ memories we add $N$ memory neurons and $2000N$
synapses (for input and output).  We also may use a relatively few
extra interneurons and synapses to implement winner-take-all dynamics.

The number of synapses is double our minimum estimate, because we
treat input and output semantics separately: they could be different.
This is much more efficient than the distributed model.  There is also
an extra neurons for each memory.  But we can readily increase the
capacity for stored memories by increasing the number of memory
neurons.  With $N=500$, we would have the same number of synapses as
in Hopfield's model, $50\%$ more neurons, but double the capacity.  

With the distributed model, the capacity can be increased only either
by simply duplicating the system, which is always a possibility, or by
a change in architecture.  An appropriate change in architecture would
be to use the original input neurons to feed a separate layer which
codes the same information more sparsely, after the style of a support
vector machine.

Note that the BMW model reliably performs the same pattern completion
task as Hopfield's model.  That is, a stimulus consisting of a few
values is completed to the values of all categories for the
corresponding memory.  The BMW model essentially performs a comparison
of the active bits in the sparse input pattern with the on-bits in the
stored pattern.  Because of the $5\%$ sparseness of the input, the
probability that an on-bit in the stimulus coincides with an on-bit in
an unrelated stored pattern is also $5\%$.  Once more than a few bits
are examined, the probability of a chance coincidence is extremely
small, with a correspondingly small misidentification probability.

\section{Statistics of neural responses}

\subsection{Two-population property of GM systems}

We have seen that the efficiency argument against GM cells disappears,
especially for semantically rich objects, like most people's
grandmothers.  But the efficiency analysis for the input
representation shows that the GM cell population is necessarily
accompanied by a population of cells carrying a distributed code.  

Experimental characterizations of the different kinds of coding can be
made by measuring the sparsity of neural responses to stimuli.

By sparsity we mean, for each cell, the fraction of stimuli to which
it responds.  We assume that, as in  \citet{GMC},
some threshold criterion is defined cell-by-cell as to whether a cell
responds or not.  Thus the neuron is treated as binary.  Other
definitions involving analog firing rates are possible, but we will
not use them.  Note that with our definition and when all cells have
the same sparsity, both the population and the lifetime sparsity are
equal, unlike the case \citep{Willmore.Tolhurst} with other definitions
of sparsity.

For a system with fully distributed coding, we expect to measure
sparsities characteristic of the input and output representations.
For example, in Hopfield's model, the sparsity is exactly $5\%$.  More
realistically, there will be a range of sparsity.

The input and output cells of a GM system will have naturally have
similar sparsities to those of all the cells in a distributed-memory
system.  But the GM cells must respond much more rarely.  So with a GM
system we expect there to be two very different populations of cells
distinguished by one population having a dramatically smaller sparsity
than the other.  Whether or not the two populations are in the same
area of the brain is not determined by general arguments.  But we will
find that, in fact, the putative GM cells of \citet{GMC} do have an
accompanying distributed-code population.  For our purposes it will be
unimportant whether the detected distributed-code population is the
one that provides the actual input and output for the detected GM-like
cells.

Expectations for the sparsity of GM cells can be provided in terms of
the repertoire size of a system, which has a connection to behavioral
data. 

There are two somewhat different kinds of memory system we will
consider.  One is typified by face recognition, where a recognized
input is categorized into one of $R$ categories, corresponding to
distinct persons; recognition is exclusive between categories.  Then
for a random sample of faces in the repertoire of the system the
sparsity of the GM cells is $1/R$, if we assume that each person is
allocated the same number of cells. 

The second case is for declarative memory (episodes, facts, etc).
Recall is by a stimulus containing a few components of the original
memory.  Since the same components could be part of other memories,
the pattern of recognition firing need not be exclusive between
memories.  Let us regard these patterns as priming the recall of the
memories.  Full conscious recall of one particular memory requires
some extra cues and modulation.  With a GM system, non-exclusive
priming recall would be at a relatively low level of firing above some
threshold, with full recall involving exclusive firing at a much
higher level.

In a memory system, a typical stimulus can evoke multiple memories.
If we let $n_{\text{m}}$ typify the number of memories evoked, then a
given GM neuron is caused to fire by a fraction $n_{\text{m}}/R$ of
stimuli.  It is therefore convenient to define an effective repertoire
size $R_{\text{eff}}=R/n_{\text{m}}$, so that the typical sparsity is
$1/R_{\text{eff}}$.  

From standard psychological data, we envisage that $R_{\text{eff}}$ is
thousands to at least millions for interesting cases.

In contrast, the distributed-code cells fire much more frequently;
this is known from data, and is necessary in order that this
population can represent a sufficiently large number of stimuli.

\subsection{General form of distribution of neural responses}

Measurements of single-cell responses concern only a small fraction of
cells and of all possible stimuli.  So we will treat data as being
from a sample over cells and stimuli, and deduce properties of the
whole system: e.g., the relative sizes of the cell populations and
their sparsities, and hence the number of categories coded for by the
GM population.  In doing this, we will quantify, and hence compensate
for, the strong biases against detection of GM cells.

Suppose we present a sample of $p$ stimuli that are randomly chosen
from some broad class (e.g., pictures of famous people, images
concerning movies that the subject has watched, pictures of
buildings).  Any particular cell $i$ responds to some fraction of
these, called the cell's (lifetime) sparsity $\alpha_i$.  The number $n_i$
of stimuli that evoke a response by the cell is taken from a binomial
distribution of mean $\alpha_in_i$:
\begin{equation}
  \label{eq:single.cell}
  P(n,\text{cell}_i)
  = \alpha_i^n (1-\alpha_i)^{p-n} \frac{ p! }{ n! \, (p-n)! }.
\end{equation}
This simply corresponds to the probability that $n$ of the stimuli are
in the response-causing class and $p-n$ are in the
non-response-causing class, as regards cell number $i$.  These two
classes of stimuli form fractions $\alpha_i$ and $1-\alpha_i$ of the whole set
of stimuli.  

We now consider a sample of cells, thereby sampling the distribution
of sparsity over cells, $D(\alpha)$.  This means that the fraction of cells
with sparsity $\alpha$ to $\alpha+d\alpha$ is $D(\alpha)d\alpha$.  Then the probability of
getting $n$ responses to the $p$ stimuli in some random chosen cell is
obtained by integrating the single cell response with the sparsity
distribution:
\begin{equation}
  \label{eq:binomial.combo}
    P(n|p)
    = \int_0^1 d\alpha \, D(\alpha) \,
      \alpha^n (1-\alpha)^{p-n} \frac{ p! }{ n!\, (p-n)! }
\end{equation}
This is a general result.  The only necessary assumption is that the
cells are randomly chosen out of some more global set of neurons
(e.g., hippocampus) and that the stimuli are randomly chosen out
of some global class.  

The value of $\alpha$ for a cell and the distribution $D(\alpha)$ depend both on
the choice of stimulus class and on the choice of the threshold for a
response.  Changing either will naturally affect the distribution.
For the data we analyze, the response criterion is given in
\citet{GMC}.

If multiple sessions and multiple subjects are considered, Eq.\
(\ref{eq:binomial.combo}) continues to apply, with $D(\alpha)$ being the
distribution averaged over subjects.  So this form is amenable for the
analysis of aggregated data.  

Observe also that the derivation of the formula does not require any
assumption about the independence of the firing of different neurons:
the formula is simply an average over all neurons in whatever area is
being sampled.  This allows the formula to be completely general, in
contrast to the model of \citet{Waydo}, which requires
that neuron-neuron correlations be neglected.

A common ansatz, as in \citet{Waydo}, is to assume a fixed sparsity
$a_1$, i.e., to set $D(\alpha)=\delta(\alpha-a_1)$.  Such a model we term a
single-population model.  But for an analysis of a possible GM
population, we must allow for at least two populations.

From a mathematical point of view, Eq.\ (\ref{eq:binomial.combo})
expands $P(n)$ in basis functions, with expansion coefficients $D(\alpha)$.
The significance to its use is four fold: (1) It relates the
distributions for different pattern numbers $p$ via a common set of
expansion coefficients $D(\alpha)$.  (2) The expansion coefficients are
non-negative.  (3) For a distributed-code population to represent all
possible stimuli, efficiency is important, i.e, using the minimum of
neurons.  This will tend to maximize the sparsity subject to other
constraints like keeping relatively low the metabolic costs
\citep{metabolism} of action-potential generation.  Thus we should
expect the sparsity of the distributed-code population to vary over a
fairly narrow range.  (4) Any GM population has an extremely small
sparsity, so that it populates just the bins with $n=0$ and $n=1$
responses to the $p$ stimuli.

\begin{figure}
  \centering
  \psfrag{alpha}{$\alpha$}
  \psfrag{D}{\hspace*{-3mm}$D(\alpha)$}
  \begin{tabular}{c@{~~}c}
     \includegraphics{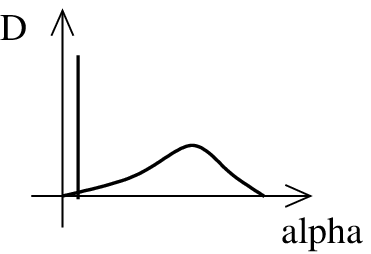}
     &
     \includegraphics{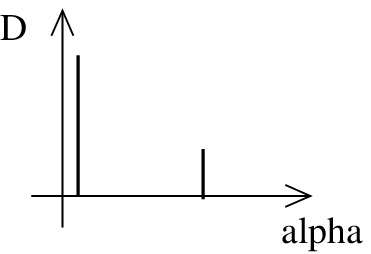}
    \\
    (a) & (b)
  \end{tabular}
  \caption{(a) Qualitative expectation for distribution of sparsity
    over cells with a GM population and a distributed-code population.
    (b) Idealization with fixed sparsity for distributed-code
    population.  For a GM population with a large repertoire, its
    sparsity is too close to $\alpha=0$ for the left hand peak to be be
    shown correctly scaled on these graphs. }
  \label{fig:D}
\end{figure}

Therefore the two population property leads to the qualitative
expectation for $D(\alpha)$ that is shown in Fig.\ \ref{fig:D}(a).  The
distribution of responses $P(n)$ can be regarded as a smeared version
of the sparsity distribution $D(\alpha)$ with $\alpha=n/p$.  Given this smearing,
a useful approximation is to replace the distributed-code peak by a
delta function at some fixed typical sparsity, Fig.\
\ref{fig:D}(b).

\subsection{Useful approximations}

Although numerical work can always be done with the linear combination
of binomial distributions Eq.\ (\ref{eq:binomial.combo}), we find it
convenient to use one of two approximations.  First, they allow simple
analytic calculations, with a consequent ease of understanding what
features of the data are important in determining particular
parameters in a model of the sparsity distribution $D(\alpha)$.  Second,
they also exhibit that for sufficiently small sparsity there is a
degeneracy in fitting $D(\alpha)$: only certain combinations of model
parameters are determined.  We verify that whenever we use these
approximations in our fits, they agree sufficiently accurately with
the underlying binomial distribution.

When the sparsity is small, the binomial distribution for a cell's
responses is approximately Poisson:
\begin{equation}
  \label{eq:poisson.approx}
  P(n,\text{cell}_i)
  \simeq (p\alpha_i)^n e^{ - p\alpha_i } \frac{ 1 }{ n! }.  
\end{equation}
This is derived by the use of Stirling's approximation, and is valid
when $\alpha_i\ll1$ and $p\gg1$, which is true in all the cases we treat.  The
approximation depends only on the product $p\alpha_i$ and not on $p$ and
$\alpha_i$ separately.

When the sparsity is so small that we can neglect the probability of
getting two or more responses, we can use what we call the GM-cell
approximation:
\begin{equation}
  \label{eq:GM.approx}
  P(n,\text{cell}_i)
  \simeq
  \begin{cases}
    1- {p}/{R_{\rm eff}} & \mbox{if $n=0$},
  \\
    {p}/{R_{\rm eff}} & \mbox{if $n=1$},
  \\
    0 & \mbox{otherwise},
  \end{cases}
\end{equation}
where we have replaced the ultra-small sparsity $\alpha_i$ by $1/R_{\rm
  eff}$ to relate it to our expectations for the sparsity of GM cell
responses.

\section{Data analysis}

We now analyze the measurements by \citet{GMC}.
For each of their experimental subjects, there was first a screening
session in which a large number of disparate images were presented.
This was sufficient to detect responsive cells, but not to measure
their selectivity.  Then there was a testing session that probed the
selectivity by using many different images of the same people and
objects to which responses were found in the screening session.  We
will find useful population information from the screening session
data alone.  

It would be a big mistake to include the testing session data in our
analysis, since the images for a testing session were systematically
chosen to concern people and objects whose images evoked a response in
the previous screening session.  The distribution $D(\alpha)$ would be
different in the two sessions.  For example, suppose that in the
screening session three images of Brad Pitt, Jennifer Aniston and
Halle Berry evoked responses from three classic GM cells.  These cells
have a very small sparsity $1/R$ with respect to images of people,
with $R$ being the subject's repertoire for recognizing faces.  Then
in a testing session that uses equal numbers of images for just these
three people, each of the three GM cells would respond to one third of
the images.  Thus with respect to the new, specially chosen stimulus
class, these cells have sparsity $1/3$.

We use the following data:
\begin{itemize}
\item Recordings were made from 343 single units and 650 multi-units.
  Given the substantial number of single units, we propose that the
  multi-units on average correspond to $2.5$ neurons, to give a total
  of approximately 2000 cells, 250 in each of 8 patients.

  The number 2000 is for cells that produced some detectable signal.
  However there are many more cells that are within range of detection
  by the extracellular electrodes used but that failed to give any
  identified action potentials (Quian Quiroga, private communication
  and \citet{Waydo,Henze,Buzsaki}).  Thus we should increase the number
  of cells by some factor $K$, whose value we will estimate later.
  Then the total number of cells available for detection is $2000K$;
  any of these would have been detected if it had given action
  potentials at rates comparable to that of the actually detected
  cells.  Our numerical results will have a very simple scaling with
  $K$.

\item There were on average $p=93.9$ stimuli in each screening session.

\item A total of 132 units produced a response above threshold.
  
\item Of these, 51 were candidate GM cells, i.e., they responded to a
  single image within the screening session.

\item The remaining 81 were not so highly selective.

\item On average, the responsive units responded to $3.1\%$ of the
  presented images, i.e., to 2.9 images.

\end{itemize}
Given the low fraction of responsive units, we assume that an
above-threshold response from a multi-unit is a response from one
particular cell.

We will analyze the data with the aid of our general expansion,
(\ref{eq:binomial.combo}).  Since we wish to test compatibility with
the GM cell hypothesis, we arrange our analysis without any initial
assumption about the necessary existence of GM cells:
\begin{enumerate}
\item First we attempt to make a fit with a conventional
  distributed-code model with a fixed sparsity.  
\item When we find this fails to be a good fit, we add a second
  component of different sparsity, as a minimal model to fit the
  data. 
\item The second component turns out to have such small sparsity that
  only its responses for $n=1$ are significant.
\item This is suggestive that there are indeed cells that approximate
  GM cells.  So we reanalyze the data in terms of a model of GM-like
  cells together with a distributed-code population, so as to
  determine appropriate properties of the GM-like population.  This
  makes it easy to allow for issues like the stimuli being or not
  being in the system's repertoire.
\end{enumerate}

\subsection{Single distributed-code population}

We first try a model of a single distributed-code population with a
single sparsity $a$.  This is the model \citep{Rolls.Treves} normally
used in theoretical work on autoassociative networks.  It corresponds
to a term $f_{\rm D}\delta(\alpha-a)$ in the general formula
(\ref{eq:binomial.combo}).  Here $f_{\rm D}$ is the fraction of cells
in this population, with the remaining neurons being silent.  (The
most conventional versions, as in \citet{Waydo}, assume $f_{\rm
  D}=1$.)  In the Poisson approximation, the probability that a
particular neuron is in the distributed-code population and that it
fires in response to $n$ out of $p$ presented images is
\begin{equation}
\label{eq:D}
  P(n ~\&~ \text{D}) \simeq 
        f_{\rm D}
         \left( pa \right)^n
           e^{ -pa }
          \frac{ 1 }{ n! }.
\end{equation}

\begin{figure}
  \centering
  \includegraphics[scale=0.6]{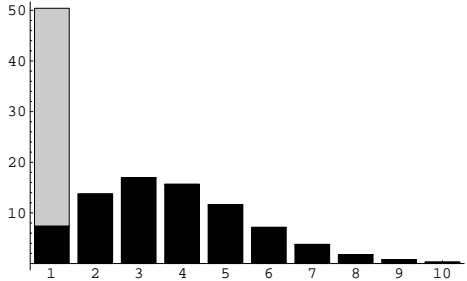}
  \caption{Distribution of cells with a particular number of
    responses. The dark bars correspond to the fitted distributed
    population and the light bar is the excess of data above the
    distributed-population contribution.  Note that the published data
    \citep{GMC} only enable us to obtain the sum of the bins with
    $n\geq2$ and the mean of $n$ restricted to these bins.  So the dark
    bars represent how cells with a distributed coding would give
    these responses.}
  \label{fig:Pn}
\end{figure}

In view of a possible GM-cell population, which would appear almost
entirely at $n=1$ and $n=0$, we fit the two parameters of the
population with data from those cells that give $n \geq 2$ responses.  In
App.\ \ref{sec:model}
we give more details of the model including its later elaboration to
include a GM population, and obtain formulae for two measurable
quantities.  One, $P(n\geq2)$, is the probability of getting 2 or more
responses from a cell; its experimental value is $81/(2000K)$.  The
second quantity is the mean value of $n$ for these cells, which we
write as $\langle n \rangle_{n\geq2}$; its experimental value is obtained from
\begin{equation}
  \langle n \rangle_{n\geq1} ~ P(n\geq1)
  ~=~ 1 \times P(1) ~+~  \langle n \rangle_{n\geq2} ~ P(n\geq2).
\end{equation}
Hence, from the data
\begin{equation}
   \langle n \rangle_{n\geq2}
=  2.9 \times \frac{132}{81} - \frac{51}{81}
=  4.1.
\end{equation}
From Eqs.\ (\ref{eq:model.ge2}) and (\ref{eq:model.mean.n.ge2})
we find that the distributed cells are a fraction
\begin{equation}
\label{eq:fD}
  f_{\rm D} = 4.6\% \frac{1}{K}
\end{equation}
of all cells, and that $pa = 3.7$.  Hence the sparsity is
\begin{equation}
\label{eq:a}
  a = \frac{3.7}{93.9} = 4\%,
\end{equation}
independently of the value of the factor $K$ for the number of silent
cells.  Given the expected statistical errors on the data, due to
finite size samples, the relative error on these numbers is about
$20\%$.  We now have a complete experimentally determined prediction
for the distribution of responses in the distributed population, the
dark bars in Fig.\ \ref{fig:Pn}.

Given this fit with rather sparsely firing cells, we find the numbers
of cells in this population that give zero and one responses:
\begin{equation}
\label{eq:extrap.D}
  \begin{tabular}{c|c|c}
          &  Frac.\ of all cells & Num.\ of cells \\
    \cline{2-3}
    $n=0$ &  $0.11\%$ & $2\pm1.5$ \\
    $n=1$ &  $0.41\%$ & $8\pm3$ \\
    $n\geq2$ &  $4.1\%$  & $81\pm9$ 
  \end{tabular}
\end{equation}
These are an extrapolation from $n\geq2$ using a natural model for sparse
distributed coding.  The errors are just from the sampling statistics:
Given the population parameters, the standard deviation of the number
of cells, $N_n$, in each bin is just $\sqrt{N_n}$.

The measured number of cells in the $n=1$ bin is 51, far in excess of
the extrapolation.  Even if there is a distribution of sparsity for
different cells in the distributed-code population, this cannot change
this deduction greatly: a cell that fires in response to at least
several percent of stimuli is likely to be detected, and relatively
few distributed-code cells give just $n=1$ and $n=0$ responses to the
$\sim100$ presented stimuli.  We also note that only around $2\%$ of the
distributed-code cells fail to get detected: these are the cells in
the $n=0$ bin.  When they are in range of the electrodes, detection of
the distributed-code cells is almost unbiased.

We deduce from the excess at $n=1$ that there is strong evidence for a
second population of ultra-sparsely firing cells, just as predicted by
general considerations if there is a GM system.  The null model, with
only a distributed population of cells together with completely silent
cells, appears to be ruled out.  Fig.\ \ref{fig:Pn} and our conclusion
about the $n=1$ excess are independent of the number of undetected
silent cells.   They are also independent of any assumption that the
cells in the second population are actual GM cells. 

Similar evidence for an excess of sparsely firing cells has
perhaps been found by \citet{Barnes}.  Their Fig.\ 8
shows an excess for some but not all hippocampal-related areas in the
rat. 

Of course, a better extrapolation could be made if experimental values
of $P(n)$ as a function of $n$ were available.  We could imagine
several populations of input cells, activated by different kinds of
image, so an improved model is a combination of several distributions
of the form (\ref{eq:D}), as in Eq.\ (\ref{eq:binomial.combo}).

\begin{figure}
  \centering
  \includegraphics[scale=0.6]{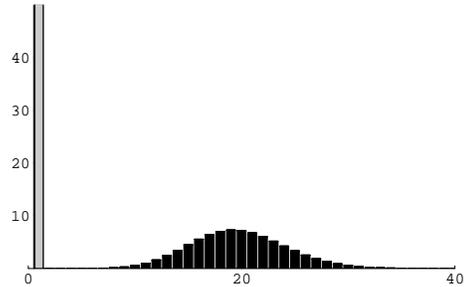}
  \caption{The predicted distribution of firing of cells when $p=500$
    stimuli are used with the same model parameters as in Fig.\
    \ref{fig:Pn}.  The distribution is now bimodal, with a clear gap
    between the distributed population response and the GM response,
    now off-scale, at $n=1$.}
  \label{fig:500}
\end{figure}

It has been said that in order to deduce the two population property
from a plot such as Fig.\ \ref{fig:Pn}, the distribution must
necessarily be bimodal.  Obviously, if we have a bimodal distribution,
the inference would be cleanest and without theoretical prejudice.  We
illustrate this in Fig.\ \ref{fig:500}, where we apply a
two-population model described below to predict the responses to
$p=500$ stimuli.  There is a GM-cell response that remains at $n=1$.
With a more limited set of stimuli, we need a theoretical expectation
for the distributed-code population to extrapolate to $n=1$ from the
data, so as to quantify a possible excess at $n=1$.  Note however that
if the distributed-code population had a distribution of sparsity
rather than one fixed sparsity, the peak in Fig.\ \ref{fig:500} would
be spread out.

\subsection{Initial two-population model}

To fit the data we need a minimum of one more population of cells,
evidently of much lower sparsity than the first population.  Without
making any initial hypothesis about its GM-like nature we try the
following ansatz for the sparsity distribution:
\begin{equation}
  \label{eq:model2}
  D(\alpha) = (1-f_{\rm D}) \, \delta(\alpha-a') + f_{\rm D} \, \delta(\alpha-a).
\end{equation}
where we preserve notation from the previous section, and resolve the
ambiguity between exchanging the definitions of the two populations by
requiring $a'<a$.  This model has three parameters to be fit to three
available measured quantities: the fraction of cells with 1 response,
the fraction of cells with 2 or more responses, and the mean number of
responses.  Within the Poisson approximation, which is always good,
and with the restriction to the 2000 detected cells, we need to solve
the following equations
\begin{align}
  \frac{N_1}{N} & = (1-f_{\rm D}) xe^{-x} + f_{\rm D} ye^{-y},
\\
  \frac{N_1+N_{\geq2}}{N} & = 
      1 - (1-f_{\rm D}) e^{-x} - f_{\rm D} e^{-y},
\\
  \frac{ \langle n \rangle_{n\geq1} \,(N_1+N_{\geq2}) }{N} & = 
      (1-f_{\rm D}) x + f_{\rm D} y.
\end{align}
From the data, we use $N=2000$, $N_1 = 51$, $N_{\geq2}=81$, and $\langle n \rangle_{n\geq1}
= 2.9$. The fit parameters are $f_{\rm D}$, and the combinations
$x=pa'$ and $y=pa$.  From this we find
\begin{equation}
  x = 0.0224, \qquad y = 3.717, \qquad f_{\rm D} =0.95, 
\end{equation}
so that the sparsities are
\begin{equation}
  a' = 2.3\times10^{-4}, \qquad a = 0.039.
\end{equation}
The sparsity of the extra population is so low that it populates only
the $n=1$ (and $n=0$) bins to a good approximation, even though it
concerns $95\%$ of the neurons.  Thus the properties of the higher
sparsity population are essentially unchanged from the
single-population fit, which confirms our original choice to fit it to
the data concerning cells with 2 or more responses to stimuli.

It now becomes useful to analyze the lower sparsity population in
terms of a GM-cell approximation (\ref{eq:GM.approx}).  This will
give us a simple way of determining whether the population's
properties are appropriate for true GM cells.  Very importantly, it
will also give us a simple way of treating certain variations in the
population's properties that are appropriate in the GM-cell context,
and of allowing for the silent-cell correction factor $K$.

\subsection{Detailed two-population model}

\begin{figure}
  \centering
  \includegraphics[scale=0.85]{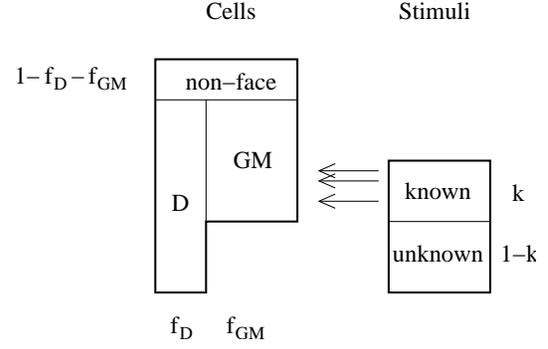}
  \caption{Classes of neurons and stimuli in a simple model with
    GM-like cells. A fraction $f_{\rm GM}$ of the facially-responsive
    cells are GM-like.  A second fraction, $f_{\rm D}$, fire in a
    distributed and less selective fashion.  The remaining cells do
    not respond to any facial stimuli.  A fraction $k$ of the stimuli
    are in the repertoire of the GM cells.}
  \label{fig:GM-model}
\end{figure}

We therefore model the responses of the cells by a population that
uses a conventional distributed code supplemented by a possible
GM-cell population, as illustrated in Fig.\ \ref{fig:GM-model}, these
comprising fractions $f_{\rm D}$ and $f_{\rm GM}$ of the total number
of cells.  The remaining cells do not respond to any stimuli at all in
the class used (which we label as ``faces'', even though some stimuli
used in \citet{GMC} were of other kinds); this fraction $1-f_{\rm
  D}-f_{\rm GM}$ is completely silent for the purposes of the
experiment.  We let $k$ be the fraction of the images used which
correspond to a memory in the GM population.

Our model corresponds to a case of the general expansion Eq.\
(\ref{eq:binomial.combo}), in which we use just two delta-functions in
$D(\alpha)$, as in Fig.\ \ref{fig:D}, with the GM-cell approximation
(\ref{eq:GM.approx}) used for the low sparsity population.  To this we
add a possible population of absolutely silent cells, i.e., a term in
$D(\alpha)$ at exactly $\alpha=0$.

As explained earlier, we let $R_{\text{eff}}$ be the effective size of
the population's repertoire, i.e., the repertoire $R$ divided by the
number of simultaneously evoked memories.  For classic GM cells
$R_{\text{eff}}=R$, of course.  Then a randomly chosen GM cell has a
sparsity $k/R_{\text{eff}}$, i.e., the probability for the image to be
in the system's repertoire times the probability to respond to an
image in the repertoire.  When $2000K$ cells are each presented with
(an average of) $p=93.9$ images of different people or objects, the
expected number of GM-like responses is therefore $2000\times K \times 93.9 \times
f_{\rm GM} \times k/R_{\text{eff}}$.  This number we found to be 43.  In
fact the choice of images was made after interviews with the subjects
(Quian Quiroga, private communication), to put them in the repertoires
of the subjects, so we now set $k=1$, to find
\begin{equation}
  R_{\text{eff}} \simeq 4400 f_{\rm GM} K.
\end{equation}
Now $f_{\text{GM}}$ is less than $1-f_{\rm D} = 0.95\mbox{~to~}1$,
depending on the value of $K$.  It is useful to define $\hat{R} =
R_{\text{eff}} /(Kf_{\text{GM}})$, which is the combination of
parameters actually determined by data.  Then $R_{\text{eff}} <
(1-f_{\text{D}})\hat{R} = 4400 K - 200$.

If we ignored the silent-cell issue and set $K=1$ we would find that
at most 4200 categories are coded for.  Now the detected cells are in
areas like the hippocampus that obviously perform additional functions
besides face recognition.  Also, the repertoire of humans for faces
and many other categories is very much larger, as measured
behaviorally \citep{Dudai}.  This would appear to imply that the
detected GM-like cells are not classic GM cells.  However:
\begin{itemize}
\item The hippocampus is not the ultimate store of long term memory,
  so that the repertoire should be perhaps only a year's worth.
\item The fraction of GM cells, $f_{\text{GM}}$ for the chosen
  stimulus class could be less than $1-f_{\text{D}}$.  The remaining
  cells would not respond to any stimuli in the class (e.g., images of
  faces); they might be distributed-code cells or GM cells for other
  classes of stimuli.  Extrapolating results on images of faces to
  other stimuli is sensible, so we would expect the vast majority of
  cells, up to a few percent corrections to be in whatever GM-like
  population is appropriate, so that the upper limit on $R_{\rm eff}$
  found above is appropriate to be applied to the whole GM population
  and not just those of the facially-responsive kind.
\item Very familiar people might have more neurons allocated to them,
  as required by the storage requirements, and some of these could be
  GM cells.  So the assumption of equal GM-cell populations for each
  category could easily be false, so that the estimate of categories
  (e.g., $4200K$) is only about those categories which are highly
  familiar to the subjects, which brings improved plausibility.
\item There could be an experimental bias in cell selection.  We made
  a basic assumption that the cells probed were a random sample in
  these areas.  But the properties of epilepsy and the selection of
  suitable subjects for study might be such that the electrodes are in
  areas preferentially responsive for visual images of people.  In
  that case, the category count refers to a restricted set of stimuli,
  improving the plausibility.
\item An interesting possibility is that the cells are not classic GM
  cells but code for memories in a GM style, as in the model 
  of \citet{BMW}.  This is of course a natural
  hypothesis for the hippocampus, if we grant the GM cell idea at all.

  Then $1/R_{\rm eff}$, as estimated above, is the average fraction of
  recent memories that concern the people pictured in the images.
  $R_{\rm eff}$ will be biased to relatively low values since the
  images were chosen to be of people well-known to the subjects.
\end{itemize}

If the cells were classic GM-cells, we can estimate the number of
cells per exclusive category.  There are roughly $10^7$ cells in one
region of the human hippocampus.  Dividing by $4200K$ gives
the number of cells per category, about $2500/K$.  With $K=1$ this
appears rather large.  But with the large value of $K$ that we will
calculate later, the number is quite modest.

Finally, from the fraction $f_{\rm D} = 4.6\%/K$ of non-GM cells, we
deduce that they number about $5\times10^5/K$.  An undoubtedly
excessively simple idea is to identify them with the input
representation.  This number could obviously much higher than the few
hundred that are perhaps needed at a minimum for coding features
relevant for face identification.  But there must obviously be many
other specialized and less specialized kinds of distributed input
representation, as well as distributed output representations.

Beyond approximating the general sparsity distribution $D(\alpha)$ by a sum
of a small number delta functions, our calculations used the Poisson
approximation for the distributed-code population and the GM
approximation for the ultra-low sparsity population.  We have verified
that using the full binomial distribution does not significantly
affect the results.

\subsection{Silent-cell correction}

The experimental estimates of the numbers of single and multiple units
in \citet{GMC} depended on detection of spikes from the neurons, even
if the spike numbers never passed the limit for the defined threshold
for a responsive neuron.  However, within detection range of the
extracellular electrodes used are many cells that never give a
detected signal.  \citet{Waydo} state that in the data
1--5 units are identified per electrode and they cite \citet{Henze} for
an estimate that that 120--140 neurons are within detection range.  To
get a first estimate we can say that on average 2 units are detected
per electrode out of 130 possible neurons, so that the neuronal
population is about 65 times the number of units.

We already estimated that about the number of neurons detected in
\citet{GMC} was about twice the number of units, to give a total of
about 2000 detected neurons.  So we should further multiply the number
of neurons by $K = 65/2 \simeq 30$, for a total of $60\,000$ neurons in
range of detectability.

This increase does not affect our estimate of the sparsity of the
response of the distributed-code neurons; that stays at $4\%$.  It also
does not affect our estimate of the \emph{relative} numbers of
\emph{detected} cells in our two populations (distributed-code v.\
GM-like).  But it does drastically decrease the fraction of
the distributed-code population to $4.6\%/K \simeq 0.015\%$.  

Most importantly it increases our estimate of the number of categories
coded.  The basic quantity here is $R_{\rm eff} \simeq 4400K \simeq 10^5$.
Given the roughness of our calculations, this is probably accurate to
a factor of 2 or so.  The key issues concern the orders of magnitude.

\section{Comparison with previous measurements of distributed
  representations}

\subsection{Waydo et al.}

\citet{Waydo} work with data that is evidently a superset
of the data that the same group published in \citet{GMC} and to which
we made a two-population fit.  They make a fit with a one-population
model that is the same as ours in the special case that all detected
cells are in the distributed-code population: $f_{\text{GM}}=0$,
$f_{\text{D}}=1$.  One superficial difference is that they use the
exact binomial distribution instead of the analytically more tractable
Poisson approximation; this makes a negligible difference for the
small sparsities in question.  Another difference is that they
normalize to units rather than neurons; but this only results in
trivial scalings of certain parameters.

\begin{table*}
  \centering
  \begin{tabular}{l l||c|c|c||c|c|c|c}
    \multicolumn{2}{l||}{$N=42$ units}
    &  \multicolumn{3}{c||}{\citet{Waydo}}
                 & \multicolumn{3}{c}{Our fit}
  \\
     &&&&&& Best &
  \\
    \multicolumn{2}{l||}{$S=88$ stimuli}
     & Data & $a_1=0.23\%$ & $a_1=0.54\%$ 
            & 84 neurons & 105 neurons & 126 neurons
  \\
  \hline
    $N_{\text{r}}$ & Responsive units
       & 7.9  & 7.7 & 15.9 & 5.5 & \bf 6.9 & 8.3
  \\
    $S_{\text{r}}$ & Evocative stimuli
       & 16.4 & 8.1 & 17.9 & 14 & \bf 17 & 20
  \\
  \hline\hline
    $S_{\text{r},\,n_{\text{r}}\geq2}$ & Fraction of stimuli with $N_{\text{r}}\geq2$
         & 4.1\% & 0.44\% & 2.2\% 
         & 1.4\% & \bf 2.1\% & 2.9\% 
  \end{tabular}
  \caption{Comparison of the per session data in \citet{Waydo}
    with the results of their one-population model, and
    with the predictions of our two-population model, using its
    already-determined population parameters.  In the two-population
    model, the number of neurons corresponding to 42 units is
    adjustable, with the best result in the middle column.
    The data as well as the fit in the column headed $a_1=0.54\%$
    are those in \citet{Waydo}. 
    For comments on the predictions in the last line, see the text. 
  }
  \label{table:comparison.session}
\end{table*}

They performed a Bayesian fit to the session-by-session data on the
joint probability of measuring $N_{\text{r}}$ responsive units and
$S_{\text{r}}$ evocative stimuli in a set of $N$ units with $S$
presented stimuli.  The probabilities in the model can be derived from
the underlying probability distribution (\ref{eq:binomial.combo}), with
one further critical assumption, that firing in the different neurons
is independent.

As can be seen from the on-line supplementary material for
\citet{Waydo}, not only is is quite difficult to derive the
distribution from the underlying distribution for the response of one
neuron, but the resulting formula involves a delicate cancellation of
opposite-sign terms.  The formula therefore needs extreme care in
numerical work.  However certain averaged quantities considered in
\citet{Waydo} are easier to derive, and in App.\ \ref{sec:evocative} we
present derivations that apply also to our two-population model, given
only the extra assumption of independence of different neurons.

The main result of the analysis in \citet{Waydo} is a distribution for
sparsity, which is to be interpreted as a posterior distribution in
the Bayesian sense for the \emph{single} sparsity $a_1$ of the neural
population.  We use the symbol $a_1$ for the sparsity to avoid
confusion with the sparsity parameter of the distributed-code
population in our fits.  When the same criterion for a neural response
as in \citet{GMC} is used, the peak of the distribution is at
$a_1=0.23\%$, which is therefore the best fit according to the usual
maximum likelihood criterion.  The distribution has a long asymmetric
tail to large $a_1$ and the average value of the posterior
distribution, at $a_1=0.54\%$, is also a useful estimate of the value
of $a_1$.

These values are much lower than the value of $a$ in our
two-population model, as is natural if the fit is to be a compromise
between matching an ultra-low sparsity GM-like population and a higher
sparsity distributed-code population.  We see this quantitatively in
Table \ref{table:comparison.session}, where we show data and the
results of their model fit and the \emph{predictions} of our model.
The data are from \citet{Waydo} and are an average over 34 sessions.
One datum is the average number of units $N_{\text{r}}$ in \emph{one}
session that respond to at least one stimulus, out of an average of
$S=88$ stimuli.  Another datum is the average number of stimuli
$S_{\text{r}}$ to which at least one unit responds in a session, out
of an average of $N=42$ detected.

The one-population model evidently has a choice, to have a relatively
low sparsity to get the correct number of responsive units, or to have
a relatively high sparsity to get the correct number of evocative
stimuli.  Our two-population model does considerably better.  We can
improve its results by increasing the number of neurons a bit more
relative to the number of units than we originally supposed.  Note
that from the first line of the table, the fraction of responsive
units is measured to be $7.9/42=19\%$.  This is substantially higher
than the measurement for the same fraction $132/993=13\%$ given in
\citet{GMC}.  So the data are not completely consistent.

\begin{table*}
  \centering
  \begin{tabular}{l||c||c|c||c}
    $p=93.9$ stimuli & \citet{GMC}
                 &  \multicolumn{2}{c||}{\citet{Waydo}}
                 &  Our fit
  \\ 
     & & \multicolumn{2}{c||}{ $N=993$ units}
  \\
              & Data & $a_1=0.23\%$ & $a_1=0.54\%$  
  \\
  \hline
    Responsive neurons with $n=1$ &  51 & 174 & 305 & Exact
  \\
    Responsive neurons with $n\geq2$ &  81 & 20  & 92 & Exact
  \\
    $\langle n\rangle_{n\geq2}$ &                     4.1 & 2.1 & 2.2 & Exact
  \end{tabular}
  \caption{Comparison of results of \citet{GMC} with the predictions of
    the one-population model of \citet{Waydo}.  Note that the three
    parameters of our 
    two-population model are determined from the three data values in
    this table. 
  }
  \label{table:comparison.global}
\end{table*}

A final piece of session-averaged data given in \citet{Waydo} is the
fraction of stimuli that produced a (simultaneous) response in at
least two neurons.  In in App.\ \ref{sec:evocative}, we derive a
formula for this quantity.  As with the number of of evocative
stimuli, a neglect of neuron-neuron correlations is needed.  The
results are shown in the last line of Table
\ref{table:comparison.session}.  As already observed in \citet{Waydo},
the one-population model with their preferred value $a_1=0.54\%$ gives
a fraction $2.2\%$ that is rather below the data ($4.1\%$).  A
comparably bad fit is obtained by our two-population model.

In fact, the bad fit happens quite generally.  The formulae for both
$S_{\text{r}}$ and $S_{\text{r},\,n_{\text{r}}\geq2}$ given them in terms
of a single property of the model, the cell-averaged sparsity
$\bar{\alpha}$.  We show in App.\ \ref{sec:evocative}, that when $\bar{\alpha}N$
is not two large, the two quantities obey an approximate relation
\begin{equation}
  \frac{ S_{\text{r},\,n_{\text{r}}\geq2} }{ S }  
  \simeq  \frac{1}{2} \left( \frac{ S_{\text{r}} }{ S } \right)^2.  
\end{equation}
This relation is obeyed to useful accuracy in the model calculations
in the last two lines in Table \ref{table:comparison.session}, but it
is violated by a factor of two by the data. The derivation is not
affected by adding yet more populations of different sparsities, but
only by including neuron-neuron correlations.

There are in fact two simple ways to overcome this problem.  One is
simply that one fraction is proportional to the number of detected
units in a session, and the other is proportional to its square.
Since this number varied quite widely \citep{Waydo} between sessions
(18 to 74), the session average of $N^2$ cannot be replaced by the
square of the average of $N$.  This could easily account for the
factor of two mismatch.  In contrast, the fraction of evocative
stimuli is amenable to a simple average over sessions.

The second possibility is from neuron-neuron correlations, to which
$S_{\text{r},\,n_{\text{r}}\geq2}$ is much more sensitive than the other
observables.  If there were a small fraction of nearby neurons that
always fired in pairs, these would disproportionately contribute to
$S_{\text{r},\,n_{\text{r}}\geq2}$, but not nearly as much to
$S_{\text{r}}$.

Of course, both these suggestions can be tested by a closer
examination of the data.

We also examine how well the one-population model, with the parameters
from \citet{Waydo}, agrees with the data we used from the earlier paper
\citep{GMC}.  This is shown in Table \ref{table:comparison.global}.  It
can be seen that the observables we used are particularly sensitive to
the differences between the models.  A low \emph{average} sparsity is
necessary to keep the number of responsive neurons down to the
experimental value.  But with a one-population model this also implies
that neurons with $n\geq2$ responses are many fewer than those with $n=1$
responses.  Moreover the number of cells with even more responses than
2 is minute, so that $\langle n\rangle_{n\geq2}$ is close to its minimum value of 2,
whereas the data is much higher.  This is a clear indication of the
need for two populations of cells with very different sparsities.

We conclude that working with the distribution of the number of
responses by individual neurons (or units), as we do, is preferable to
the other distributions.  The distribution $P(n|p)$ is easy to work
with, and data can be usefully aggregated over a whole experiment,
given only that the number of stimuli is approximately the same in
each session and that the stimuli are chosen at random in some large
class.  These conditions can be imposed by an experimenter.  The model
can always be systematically improved by changing the sparsity
distribution, e.g., by adding extra components.  Working with
aggregate data keeps the sampling errors usefully low, and parameters
for a model can be computed simply from properties of the aggregate
data.

\subsection{Abbott, Rolls, and Tovee}

Other analyses of data, e.g., 
\citet{Abbott.Rolls.Tovee}, have reported that hippocampal facially
responsive cells carry a distributed code as opposed to a GM-type
code.  See also the recent work of \citet{HKPC}.  These
analyses might appear to contradict our calculation that
distributed-code cells are a very small fraction, perhaps less than
$0.2\%$ of the total.  However, there is a strong bias against actually
detecting GM-like cells.  In this section, we use our fit to the more
recent data to quantify this bias, at least roughly, to determine
whether there is consistency between our results and the earlier data.

The primary issue is that experiments typically only report those
cells that are actually detected to respond to at least one of the
stimuli used.  For example, in a paper documenting place cell
\citet{silent.cells} state ``the electrode
assemblies were advanced until one or more hippocampal complex-spike
cells were isolated extracellularly.''  Then they observe that cells
that do not produce any detectable spikes ``are excluded from analysis
here due to our lack of ability to detect them''.  Since GM-like cells
respond to a very small fraction of stimuli, the ones that respond to
no presented stimulus, i.e., the vast majority, are typically omitted
from an analysis.

The resulting bias can be seen in the data that we analyzed earlier.
A minority of the detected cells in \citet{GMC} are in the GM-like
class (43 out of 132), even though we have shown that the GM-like
cells can be in the vast majority ($99\%$ or more).

With fewer stimuli, the bias becomes even stronger, as in the data
used by \citet{Abbott.Rolls.Tovee}.  They used
20 face stimuli, and the total number of facially responsive neurons
was 14.  A rather higher sparsity was reported than our result.  But
this is partly because a different definition of sparsity was used,
applied to the spike numbers rather than to a binary response
criterion.  Furthermore the cells have considerably larger background
firing rates than those in the new data.

Despite the differences in cells and species, we blindly apply our
model to give a rough test of consistency.  In our two-population
model, the fraction of GM cells in the \emph{detected} cells is
\begin{equation}
  \frac{ p / \hat{R} }
       { p / \hat{R} + f_{\rm D} \left( 1- e^{-pa} \right)
       }
= \frac{1}{1 +  \dfrac{202}{p} (1-e^{-0.04p}) }.
\end{equation}
This is the probability that a cell is a GM-like cell
\emph{conditional} on the cell producing a detected response to one or
more of $p$ stimuli.  Notice that the silent-cell ratio $K$ cancels in
this formula; we have a relation between numbers of different kinds of
detected cell under different experimental conditions.  We have
estimates for the parameters of the model, so substituting $p=20$
predicts a detected GM-cell fraction of $0.15$, i.e., about 2 cells,
in the set of cells investigated in \citet{Abbott.Rolls.Tovee}.  In
fact, \citet{Abbott.Rolls.Tovee} did report that two of their cells had
a GM-like response.  There is, of course, no significance to the fact
that this number is exactly the value predicted: there are expected
statistical fluctuations, and the measurements were done with
different methods and in a different species than in \citet{GMC}.

Nevertheless it is very important that the previous report, viewed as
evidence in favor of distributed representations, is completely
compatible with GM cells being in the vast majority, with the
parameters we have determined.  The undetected GM cells simply appear
to be silent within the experiment and are therefore classed as not
facially responsive.  Statements about the neural representation being
distributed apply only to (most of) those cells that the measurements
actually detected, not to all cells in the relevant region of the
brain.

\section{Discussion}

The results of \citet{GMC} clearly suggest the
detection of grandmother cells in the classic sense.  Many other
experiments have detected individual cells with strikingly specific
responses (e.g., \citet{HVC-RA,dg.cell.classes,silent.cells}).
Therefore it is useful to hypothesize that some of these cells are
indeed GM-like cells, even though the concept of GM-cell may need to
be extended and modified.  

A purely experimental direct test of the idea needs too many stimuli
to be practical, cf.\ \citet[p.\ 179]{Churchland.Sejnowski}.  So other
arguments must be brought in, of which we have provided two.  One uses
an estimate of the actual storage requirements for a memory system.
We showed that GM systems can be optimally efficient in the use of
synapses and neurons.  The usual efficiency argument applies only to
the input representation, but now carries the implication that in a
GM-like system there must be two populations of cells with widely
different sparsities.

Our second argument is a method to analyze neural responses.  A
particular aim is to measure whether they are quantitatively
consistent there being separate neurons coding for each recognized
person, or, alternatively, for each individual declarative memory.
Our method enables one to determine whether or not individual cells
necessarily code for multiple persons or memories.  We derived a
general formula Eq.\ (\ref{eq:binomial.combo}) for the neural
responses in terms of an underlying distribution of sparsity.  Our
expansion is a new result and is applicable independently of any
detailed theory or model of neural function.

In effect, the formula enables us to extrapolate from limited data to
obtain the fraction to stimuli to which cells respond.  It also allows
us to compensate for the strong biases involved in detecting cells
when sparsities differ by very large factors.  Thus we obtain valid
estimates of the numbers of cells of different kinds.  We thereby
solve some of the issues raised by 
\citet{Olshausen.Field} concerning the publication of data only about
responsive cells.  One primary remaining bias is that different
neurons may have different electrical characteristics, with a
consequent different maximum distance from the electrodes for
detectability of spikes.  But this is presumably a milder effect than
that caused by orders of magnitude differences in sparsities.

\subsection{Two populations essential}

From the data we find indeed that the two-population property is
obeyed.  Not only does the ultra-low-sparsity population comprise the
vast majority of cells in the brain regions concerned (hippocampus,
etc), but its sparsity can be in a range compatible with the
hypothesis of a GM-like system: Roughly $10^{-3}\%$ with a repertoire
of $10^5$.

An important role is played by the many silent cells.  It is obviously
unreasonable to assume they have no function.  But on the GM-cell
hypothesis they naturally are to be interpreted as the majority of GM
cells that are not relevant to the particular stimuli used in an
experiment.  The large number of these cells is what enables one to
overcome the strong biases against detecting a response of any one GM
cell to a limited set of stimuli.

Now the group responsible for the analyzed data argue \citep{GMC,Waydo}
that their data do not support the GM cell idea.  In 
\citet{Waydo}, they say ``if we assume that a typical adult
recognizes between 10,000 and 30,000 discrete objects (Biederman,
1987), $a=0.54\%$ implies that each neuron fires in response to 50 --
150 distinct representations.''  [
$a$ should be replaced by $a_1$ in the
notation of the present paper.]

However their analysis assumed a single value of sparsity.  While this
is a suitable approximation for conventional mechanisms of distributed
memory, it is very bad for GM-like systems.  Even though the explicit
aim of \citet{Waydo} was to test the GM-cell hypothesis, the use of a
single sparsity in effect imposed an assumption that the hypothesis is
wrong.

We showed that the single-population hypothesis is a bad fit to the
data.  Since our expansion (\ref{eq:binomial.combo}) is very general,
the fault is in the single-population hypothesis not in any assumption
about neural properties.  The rather low value of sparsity given by
Waydo et al.\ is merely a compromise between the widely different
sparsities of the two populations.  Our results are consistent with an
even higher number of recognized objects than in the estimates of
\citet{Biederman}.  Indeed, even \emph{without} allowing for
the silent cell correction, our fits allow a GM-cell population with a
sparsity of $1/4200 = 0.024\%$ corresponding to a number of objects not
far from the lower edge of Biederman's range.

Note that our basic estimate of the number of categories, $10^5$,
assumes that the cells are classic GM cells, each responding to a
single individual person.  But the number of categories could be
substantially higher.  If the cells are general memory cells, in the
style of the  model of \citet{BMW}, they could
respond to images of several people.  It could also be that more
familiar stimuli, with richer associations, have more cells.  In that
case measurements with familiar stimuli, as is the case in the data,
would be biased towards these memories with unusually large numbers of
cells, with a corresponding reduction in our estimate of the number of
categories compared with the true number.

\subsection{Anatomy}

In GM systems, like the model of \citet{BMW}, the
number of GM cells is very much larger than that of the input cells,
as is consistent with the numbers we have deduced.  An immediate
implication is that each GM cell receives input from a modest number
of input cells, but that each input cell sends output to a much larger
number of memory cells.  Given also our finding that the GM cells in
the relevant regions are in the vast majority, there are some striking
anatomical implications.

In fact striking disparities in synapse number are well known in the
hippocampus \citep{Amaral}: For example, each CA3 pyramidal cell gets
about 50 input synapses from dentate granule cells, while other
connections have tens of thousands of synapses.  Note that hippocampal
neurogenesis results in dentate granule cells, highly appropriate if
they are GM-like.  However, general-purpose memories need a wider
variety of (processed) input than does a face recognition system, and
hippocampal-related regions are sufficiently complex that the real
picture is undoubtedly much more complicated.  Even so, a careful
analysis of the disparities in synapse number should provide critical
information on neural function and the viability of GM-like systems.

\subsection{Other arguments against GM systems}

Other less quantitative arguments have been advanced against the
reality of GM systems, e.g.,
\citet{Churchland.Sejnowski,Rolls.Treves}.  

For example, distributed memory systems are said to be robust against
partial destruction, since there is no single location for a single
memories.  But we do know that memories disappear.  If there are
multiple GM cells for a memory in different places, then we can
overcome the robustness argument by simple redundancy.  Moreover
memories form a network of knowledge, so that individual items of
semantic memory can be readily reconstructed from other knowledge.
Episodic memory is really an ordered sequence of individual episodes,
not necessarily remembered at all precisely.  Any one episode that
disappears can be approximately filled in from neighboring episodes.

Distributed memory systems are also said to be good at filling in
missing parts of input data, as in reconstructing a full remembered
image from a stimulus containing only a part of the image.  But this
property can also be true for GM-like systems.  For example, when the
BMW architecture \citep{BMW} is used with a sparse input representation
and suitable dynamics for its GM cells, it also performs pattern
completion; the completion property is actually associated with
properties of sparse representations used for input data.

It has been said that new memories are harder to construct in GM
systems than in distributed-memory systems.  But now that adult
neurogenesis in the hippocampus is well established, it may well be
that there is actually a pool of new neurons available for at least
some uses that could include being GM-like cells for new memories.
The new neuron rate may however be excessively small.  In addition, it
is possible that the GM nodes are on dendritic tree rather than being
whole neurons.  It is known that there can be substantial changes in
dendritic topology, which could easily include the formation of new
nodes.  Here the fundamental mode of operation is of a GM-like system
while the neural code of memory neurons takes on some of the aspects
of distributed memory.  In any case, there are potential realistic
mechanisms for the formation of new GM nodes, so that there is no
insuperable obstacle here.

\section*{Acknowledgments}
JCC is supported in part by the U.S. D.O.E\@.  He also thanks Larry
Abbott and the Volen Center at Brandeis University for hospitality
during the initial stages that led to this work.  DZJ is supported by
the Alfred P. Sloan Fellowship and by the Huck Institute of Life
Sciences at Penn State.  We thank Larry Abbott, Jayanth Banavar, Gong
Chen, Anne Graybiel, John Lisman, and Michael Wenger for useful
conversations.


\appendix

\section{Model of neural firing}
\label{sec:model}

Our model for the statistics of neural firing has two cell
populations: one that uses a conventional distributed code with a
single sparsity $a$ and a second GM-cell population, as illustrated in
Fig.\ \ref{fig:GM-model}.  These from fractions $f_{\rm D}$ and
$f_{\rm GM}$ of the total number of cells.  A remaining population of
cells does not respond to any stimuli at all in the class used in the
experiment.  We let $k$ be the fraction of the images used which have
stored representations in the GM population, we let $R$ be the
repertoire of the GM cells, and we let $n_{\text{m}}$ be the typical
number of categories (or GM groups) evoked by a stimulus in the
system's repertoire.  As before, we let
$R_{\text{eff}}=R/n_{\text{m}}$.

Suppose first we record from some random cell known to be in the
distributed population.  We have seen that when we present $p$ images,
the probability of getting $n$ responses is approximately the Poisson
distribution in Eq.\ (\ref{eq:single.cell}) with $\alpha=a$.

If, instead, we pick a GM cell, then for each individual image it has
a probability $kn_m/R$ of responding.  Therefore over a set of $p \ll
R_{\text{eff}}$ 
unrelated images it has a probability $k p/R_{\text{eff}}$ of responding exactly
once.  There is a negligible probability of $n\geq2$ for such a cell.

Finally, if the cell is outside the above two populations, it is
silent in the experiment and always gives $n=0$.

Summing over the distributions of $n$ for cells of the different
kinds, weighted by their fractional population size, gives
\begin{equation}
  P(n) \simeq
     \begin{cases}
         1 
         - \dfrac{ f_{\rm GM} \, k \, p }{ R_{\text{eff}} }
         - f_{\rm D} ( 1 - e^{ - pa } )
         & \text{if $n=0$,}
     \\[2mm]
       \dfrac{ f_{\rm GM} \, k \, p }{ R_{\text{eff}} }
       + f_{\rm D} \,pa  \, e^{ -pa}
         & \text{if $n=1$,}
     \\[2mm]
         f_{\rm D} (pa)^n e^{ - pa } \dfrac{ 1 }{ n! }
         & \text{if $n\geq2$.}
     \end{cases}
\end{equation}
This is of the form of the general distribution Eq.\
(\ref{eq:binomial.combo}) with 
\begin{align}
  \label{eq:D.model}
  D(\alpha) ={}& f_{\rm D} \, \delta(\alpha-a) 
         + f_{\rm GM} k \, \delta(\alpha-1/R_{\text{eff}})
\nonumber\\ &
         + (1-f_{\rm D} - f_{\rm GM} k) \, \delta(\alpha), 
\end{align}
and with the Poisson approximation approximation for the
distributed-code population and with the GM approximation that the
GM-like cells fire so rarely that their responses for $n\geq2$ can be
neglected.

There are several meaningful parameters for the GM population, but
only one combination affects the distribution $P(n)$.  So we define
$\hat{R} = R_{\text{eff}} / (kf_{\text{GM}}) 
= R / (n_{\text{m}}kf_{\text{GM}})$, to find
\begin{equation}
  P(n) \simeq
     \begin{cases}
         1 
         - \dfrac{ p }{ \hat{R} }
         - f_{\rm D} ( 1 - e^{ - pa } )
         & \text{if $n=0$,}
     \\[2mm]
       \dfrac{ p }{ \hat{R} }
       + f_{\rm D} \,pa  \, e^{ -pa}
         & \text{if $n=1$,}
     \\[2mm]
         f_{\rm D} (pa)^n e^{ - pa } \dfrac{ 1 }{ n! }
         & \text{if $n\geq2$.}
     \end{cases}
\end{equation}
The parameter $\hat{R}$ has the meaning that if a cell is outside the
distributed-code population then it responds to a stimulus in the
chosen global class with probability $1/[\hat{R}\, (1-f_{\text{D}})] \simeq
1/\hat{R}$, where the last approximate equality applies in the
realistic case that $f_{\text{D}}$ is small, according to our fit.

We wish to extract the properties of the distributed-code cells
without contamination from the GM cells. For that we need properties
of the distribution for $n\geq2$, for which we use the probability and
the mean number of responses.  The probability of $n\geq2$ is
\begin{equation}
\label{eq:model.ge2}
  P(n\geq2) = 
     f_{\rm D} 
     \left[
        1 - (1+pa) e^{-pa} 
     \right] .
\end{equation}
The mean number of responses, in cells with $n\geq2$, is
\begin{align}
\label{eq:model.mean.n.ge2}
  \langle n \rangle_{n\geq2}
= {}&
  \frac{ \sum_{n\geq2} n P(n) }
       { P(n\geq2) }
\nonumber\\
\simeq {}&
  \frac{ pa \, (1 - e^{-pa}) }
       { 1 - (1+pa) e^{-pa} }.
\end{align}
These last two equations suffice to determine $a$ and $f_{\rm D}$ from
the data in \citet{GMC} --- see Eqs.\ (\ref{eq:fD}) and (\ref{eq:a}).

\section{Further applications of model}
\label{sec:evocative}

Several further observables are considered by 
\citet{Waydo}.  These observables refer to a session in which $S$
stimuli are presented to $N_{\text{c}}$ cells or $N$ units.

One observable is the number of neurons $N_{\text{r}}$ that respond to
at least one stimulus.  Its average is just the number of cells or
units times the probability that one cell or unit responds, which in
our notation is $P(n\geq1|S)$, in the notation of Eq.\
(\ref{eq:binomial.combo}).  Hence the number of responsive units in
one-population model of \citet{Waydo} is
\begin{equation}
  P(n\geq1|S) \times \text{\# units} = (1-e^{-Sa_1}) N.
\end{equation}
In our two-population model it is
\begin{equation}
    P(n\geq1|S) \times N_{\text{c}}
    = \left( \frac{SK}{\hat{R}} + Kf_{\text{D}}(1-e^{-Sa}) \right)
      N_{\text{c}}.
\end{equation}
We are not quite sure how many cells correspond to each unit in the
new data, so in Table \ref{table:comparison.session} we gave results
for several choices of the ratio of cells to units, $N_{\text{c}}/N$:
2 (as we estimated for the earlier data), 2.5, and 3.

A second observable is the number $S_{\text{r}}$ of stimuli that
evoked a response in at least one neuron in the session.  To derive
this from the response distributions requires a further assumption
that correlation between the firing of different detected neurons can
be neglected.

Now the probability of one stimulus evoking no response in any of $N$
independent cells (or units) is $P(0|1)^N$, where $P(0|1)$ is the
probability of no response in one cell/unit on presentation of 1
stimulus.  Hence the average number of evocative stimuli in a session
is
\begin{equation}
  S_{\text{r}} = S \left[ 1 - P(0|1)^N \right].
\end{equation}
From Eq.\ (\ref{eq:binomial.combo}), we find that in general $P(0|1) =
1-\bar{\alpha}$, where $\bar{\alpha}$ is the sparsity averaged over cells.  In
the one-population model of \citet{Waydo} we therefore get
\begin{equation}
  S_{\text{r}} = S \left[ 1 - (1-a_1)^N \right],
\end{equation}
while in our two-population model it is 
\begin{equation}
\label{eq:S.r}
  S_{\text{r}} 
= S \left[ 1 - 
       \left( 1 - \frac{K}{\hat{R}} 
                - K f_{\text{D}} a
       \right)^{N_{\text{c}}}
    \right].
\end{equation}
The appearance of $K$ in this last formula is misleading: the
dependence on $K$ of $\hat{R}$ and $f_{\text{D}}$ in our fit cancels
the explicit factor of $K$.  We have used the number of cells
$N_{\text{c}}$ in this formula rather than the number of units $N$,
since our fit is made with respect to cells.

A final observable we consider is the number of stimuli in a session
that evoked responses in 2 or more cells/units.  This quantity,
denoted $S_{\text{r},\,n_{\text{r}}\geq2}$ can be obtained from the
distribution of the number of neurone responding to a single stimulus:
\begin{align}
  P(\mbox{$n_{\text{r}}$ for 1 stim.})
  ={}& \bar{\alpha}^{n_{\text{r}}} \, (1-\bar{\alpha})^{N-n_{\text{r}}}
    \frac{N!}{ n_{\text{r}}! \, (N-n_{\text{r}})! }
\nonumber\\
  \simeq{}& \frac{ [N \, (1-\bar{\alpha})]^{n_{\text{r}}} }{ n_{\text{r}} } e^{-N\,(1-\bar{\alpha})},
\end{align}
It follows that on average
\begin{equation}
\label{eq:S.r.nr.ge.2}
  S_{\text{r},\,n_{\text{r}}\geq2}
  = S \left[ 1 - (1-\bar{\alpha})^N - N\bar{\alpha} (1-\bar{\alpha})^{N-1} \right].
\end{equation}

From Eqs.\ (\ref{eq:S.r}) and (\ref{eq:S.r.nr.ge.2}), we get a
relation between $S_{\text{r}}$ and $S_{\text{r},\,n_{\text{r}}\geq2}$
valid when
$N\bar{\alpha}$ is less than about unity and $N$ is substantially larger
than unity.  We expand the powers of $1-\bar{\alpha}$ for small $\bar{\alpha}$
to obtain
\begin{equation}
  \frac{ S_{\text{r}} }{ S }  \simeq  N\bar{\alpha}
  \qquad \mbox{if $N\bar{\alpha} \lesssim  1$},
\end{equation}
and then
\begin{equation}
  \frac{ S_{\text{r}}(N_{\text{r}}\geq2) }{ S }  
  \simeq  \frac{ (N\bar{\alpha})^2 }{ 2 }
  \simeq  \frac{1}{2} \left( \frac{ S_{\text{r}} }{ S } \right)^2
  \qquad \mbox{if $N\bar{\alpha} \lesssim  1$}.
\end{equation}




\begin{thebibliography}{99}

\bibitem[Abbott, Rolls and Tovee(1996)]{Abbott.Rolls.Tovee}
   L.F. Abbott, E.T. Rolls, and M.J. Tovee,
   ``Representational capacity of face coding in monkeys'', Cerebral
   Cortex \textbf{6}, 498 (1996).

\bibitem[Amaral, Ishizuka and Claiborne(1970)]{Amaral}
   D.G. Amaral, N. Ishizuka, and B. Claiborne,
   ``Neurons, Numbers and the Hippocampal Network'', in 
   Progress in Brain Research vol.\ 83, ed.\ J. Storm-Mathisen,
   J. Zimmer and O.P Ottersen, pp. 1--11 (Elsevier, 1970).

\bibitem[Amit, Brunel and Tsodyks(1994)]{Amit.Brunel.Tsodyks}
   D.J. Amit, N. Brunel, and M.V. Tsodyks,
   ``Correlations of cortical Hebbian reverberations: Theory versus
   experiment'', 
   J. Neurosci.\ \textbf{14}, 6435 (1994).

\bibitem[Barnes et al.(1970)]{Barnes}
   C.A. Barnes, B.L. McNaughton, S.J.Y. Mizumori, B.W. Leonard, and
   L.-H. Lin, 
   ``Comparison of spatial and temporal characteristics of neuronal
   activity in sequential stages of hippocampal processing'', 
   Progress in Brain Research vol.\ 83, ed.\ J. Storm-Mathisen,
   J. Zimmer and O.P Ottersen, pp. 287--300 (Elsevier, 1970).

\bibitem[Baum, Moody and Wilczek(1988)]{BMW}
   E.B. Baum, J. Moody, and F. Wilczek,
   Biol.\ Cybern.\ \textbf{59}, 217--228 (1988).

\bibitem[Biederman(1987)]{Biederman}
   I. Biederman, ``Recognition-by-components: a
   theory of human image understanding'', Psychol.\ Rev.\ \textbf{94},
   115--147 (1987).

\bibitem[Buzs\'aki(2004)]{Buzsaki}
   G. Buzs\'aki,
   ``Large-scale recording of neuronal ensembles'',
   Nature Neurosci.\ \textbf{7}, 446 (2004).

\bibitem[Chklovskii, Mel and Svoboda(2004)]{CMS}
   D.B. Chklovskii, B.W. Mel, and K. Svoboda,
   ``Cortical rewiring and information storage'',
   Nature, \textbf{431}, 782 (2004).

\bibitem[Churchland and Sejnowski(1992)]{Churchland.Sejnowski}
   P.S. Churchland and T. Sejnowski,
   ``The computational brain''
   (MIT Press, Cambridge MA, 1992).

\bibitem[Dudai(1997)]{Dudai}
   Y. Dudai, 
   ``How big is human memory, or On being just useful enough'',
   Learning \& Memory, \textbf{3}, 341 (1997).

\bibitem[Gardner-Medwin and Barlow(2001)]{GMB}
   A.R. Gardner-Medwin and H.B. Barlow, 
   ``The Limits of Counting Accuracy in Distributed Neural
   Representations'',
   Neural Computation \textbf{13},477 (2001). 

\bibitem[Hahnloser, Kozhevnikov and Fee(2002)]{HVC-RA}
   R.H.R. Hahnloser, A.A. Kozhevnikov, and M.S. Fee, 
   ``An ultra-sparse code underlies the generation of neural sequences
   in a songbird'', 
   Nature \textbf{419}, 65 (2002).

\bibitem[Henze et al.(2000)]{Henze}
   D.A. Henze, Z. Borhegyi, J. Csicsvari, A. Mamiya, K.D. Harris, and
   G. Buzs\'aki,
   ``Intracellular Features Predicted by Extracellular Recordings in
   the Hippocampus In Vivo'',
   J. Neurosci.\ \textbf{84}, 390 (2000).

\bibitem[Hopfield(2006)]{Hopfield}
   J.J. Hopfield, 
   ``Searching for memories, Sudoku, implicit check-bits, and the
   iterative use of not-always-correct rapid neural computation'',
   q-bio.NC/0609006.

\bibitem[Hung et al.(2005)]{HKPC}
   C.P. Hung, G. Kreiman, T. Poggio, and J.J. DiCarlo,
   ``Fast Readout of Object Identity from Macaque Inferior Temporal
   Cortex'', 
   Science \textbf{310}, 863 (2005).

\bibitem[Jung and McNaughton(1993)]{dg.cell.classes}
   M.W. Jung and B.L. McNaughton,
   ``Spatial Selectivity of Unit Activity in the Hippocampal Granular
   Layer'', 
   Hippocampus \textbf{3}, 165 (1993).

\bibitem[Kanerva(1988)]{kanerva}
   P. Kanerva, ``Sparse distributed memory'' (MIT, 1988).

\bibitem[Konorski(1967)]{Konorski}
   J. Konorski, ``Integrative activity of the brain: an
   interdisciplinary approach'' (University of Chicago Press, 1967). 

\bibitem[Lennie(2003)]{metabolism}
  P. Lennie, ``The Cost of Cortical Computation'', Curr.\ Biol.\
  \textbf{13}, 493--497 (2003). 

\bibitem[Olshausen and Field(2005)]{Olshausen.Field}
   B.A. Olshausen and D.J. Field,
   ``How Close Are We to Understanding V1?'',
   Neural Computation \textbf{17}, 1665--1699 (2005);
   ``What is the other 85\% of V1 doing?'',
   in ``23 problems in systems neuroscience'',
   J.L. van Hemmen and T.J. Sejnowski (eds.)
   (Oxford University Press, 2006).

\bibitem[Page(2000)]{Page}
   M. Page,
   ``Connectionist modeling in psychology: A localist manifesto'',
   Behavioral and Brain Sciences \textbf{23}, 443 (2000). 

\bibitem[Quartz and Sejnowski(1997)]{Quartz.Sejnowski}
   S.R. Quartz and T.J. Sejnowski, 
   ``The neural basis of cognitive development: A constructivist
   manifesto'', 
   Behavioral and Brain Sciences \textbf{20}, 537 (1997). 

\bibitem[Quian Quiroga et al.(2005)]{GMC}
  R. Quian Quiroga, L. Reddy, G. Kreiman, C. Koch, and I. Fried,
  ``Invariant visual representation by single neurons in the human
  brain'',
  Nature \textbf{435}, 1102 (2005).

\bibitem[Rolls(2001)]{Rolls}
   E.T. Rolls,
   ``Representations in the brain'',
   Synthese \textbf{129}, 153 (2001).

\bibitem[Rolls and Treves(1998)]{Rolls.Treves}
   E.T. Rolls and A. Treves,
   ``Neural Networks and Brain Function'',
   (Oxford Univ.\ Press, 1998).

\bibitem[Stepanyants, Hof and Chklovskii(2002)]{SHC}
   A .Stepanyants, P.R. Hof, and D.B. Chklovskii,
   ``Geometry and structural plasticity of synaptic connectivity'', 
   Neuron, \textbf{34}, 275 (2002).

\bibitem[Thompson and Best(1989)]{silent.cells}
   L.T. Thompson and P.J. Best,
   ``Place Cells and Silent Cells in the Hippocampus of
   Freely-Behaving Rats'',
   J. Neurosci.\ \textbf{9}, 2382 (1989).

\bibitem[Treves and Rolls(1991)]{Treves.Rolls1}
   A. Treves and E.T. Rolls,
   ``What determines the capacity of autoassociative memories in the
   brain?'', 
   Network \textbf{2}, 371 (1991).

\bibitem[Treves and Rolls(1992)]{Treves.Rolls2}
   A. Treves and E.T. Rolls,
   ``Computational constraints suggest the need for two distinct input 
   systems to the hippocampal CA3 network'',
   Hippocampus \textbf{2}, 189 (1992).

\bibitem[Waydo et al.(2006)]{Waydo}
  S. Waydo, A. Kraskov, R Quian Quiroga, I. Fried, and C. Koch,
  ``Sparse Representation in the Human Medial Temporal Lobe'', 
  J. Neurosci.\ \textbf{26}, 10232 (2006).

\bibitem[Willmore and Tolhurst(2001)]{Willmore.Tolhurst}
   B. Willmore and D.J. Tolhurst
   ``Characterizing the sparseness of neural codes'',
   Network \textbf{12}, 255 (2001).

\end{thebibliography}
\end{document}